%% file: main.tex
\documentclass{article}
\usepackage[utf8]{inputenc}
\usepackage{stackengine}
\usepackage{amsfonts,amsmath,amssymb,amsthm}
\usepackage[letterpaper, portrait, margin=1in]{geometry}
\usepackage{verbatim,float,url,graphicx,psfrag}
\usepackage{natbib}
\usepackage{graphicx}
\usepackage{mathtools} %for better underbraces
\usepackage{cases} 
\usepackage{mathabx}   %for better underbraces
\usepackage{hyperref} % to create click-able back to specific equations
\usepackage{bbm}      % for more math symbols
\usepackage{cancel}   % for strike-out symbol
\usepackage[normalem]{ulem}  %% For strikeout
\usepackage{mathtools} % for better matrices
\usepackage{dsfont}
\usepackage{color}
\usepackage[export]{adjustbox} % http://ctan.org/pkg/adjustbox
\usepackage{tabularx} % For line breaks in tables
\usepackage{tabulary}
\usepackage{marginnote}
\usepackage{multirow}
\usepackage{sistyle}
\SIthousandsep{,}
\usepackage{authblk}
\input{macros.tex}

\usepackage{xcolor}
\usepackage{ulem}%

\title{Defining 3-dimensional marine provinces with phytoplankton compositions}
\author[1]{Rafael Catoia Pulgrossi\thanks{rcatoia@ucsc.edu}}
\author[2]{Nathan L R Williams \thanks{nathanwi@ucsc.com}}
\author[3]{Yubin Raut\thanks{yubin89@mit.edu}}
\author[2]{Jed Fuhrman\thanks{fuhrman@usc.edu}}
\author[1]{Sangwon Hyun*\thanks{sangwonh@ucsc.edu}}
\affil[1]{Department of Statistics, University of California, Santa Cruz}
\affil[2]{Department of Biological Sciences, University of Southern
  California}
\affil[3]{Earth, Atmosphere and Planetary Sciences, Massachusetts Institute of Technology}
\begin{document}
\maketitle
\newpage
\begin{abstract}
\input{abstract}
\end{abstract} 

\include{main-content}

\include{data-statement}
\include{author-contribution}
\include{supp-content}

\bibliographystyle{unsrt}
\bibliography{main}
 
\end{document}

%% file: macros.tex
\usepackage{booktabs}
\usepackage{subfig}
\usepackage{rotating} 
\usepackage{dsfont}
\usepackage[long]{optidef}

\def\R{\mathbb{R}}

\def\gh
\def\cI{\mathcal{I}}

\usepackage{pbox}

\newcommand\lat{\mathrm{lat}}
\newcommand\depth{\mathrm{depth}}

\usepackage{tikz}
\usetikzlibrary{calc,tikzmark} 
\newlength{\imageheight}

%% file: abstract.tex
Marine provinces rarely include fine-resolution biological data, and are often defined spatially across only latitude and longitude. Therefore, we aimed to determine how phytoplankton distributions define marine provinces across 3-dimensions (i.e., latitude, longitude, and depth). To do this, we developed a new algorithm called \texttt{bioprovince} which can be applied to compositional biological data. The algorithm first clusters compositional samples to identify spatially coherent groups of samples, then makes flexible province predictions in the broader 3d spatial grid based on environmental similarity.  
We applied \texttt{bioprovince} to phytoplankton Amplicon Sequencing Variants (ASVs) from five, depth-resolved ocean transects spanning north-south in the Pacific Ocean. In the surface layer of the ocean, our method agreed well with traditional Longhurst provinces. In some cases, the method revealed that with more granular taxonomic resolution afforded by ASVs, traditional Longhurst provinces were divided into smaller zones, mainly due to differences in the composition of \emph{Prochlorococcus} ecotypes. 
One of the major advances of this method is its ability to incorporate a third dimension, depth. Indeed, our analysis found significant depth-wise partitions throughout the Pacific with remarkable agreement in the equatorial region with the base of the euphotic zone. In contrast, the Antarctic Longhurst province
showed vertical homogeneity in ASV compositions, with our 3-dimensional bioprovinces extending with no division from the surface down to aphotic depth. Our algorithm's ability to delineate 3-dimensional bioprovinces will enable scientists to discover new ecological interpretations of marine phytoplankton ecology and biogeography.  Furthermore, as compositional biological data inherently exists in three spatial dimensions in nature, bioprovince is broadly applicable beyond marine plankton, offering a more holistic perspective on biological provinces across diverse environments.


%% file: main-content.tex
\section{Introduction}

Partitioning and delineating marine areas into distinct regions directly enables a better understanding of the distribution of organisms and ecosystem processes. These regions are generally made by dividing the ocean into coherent and non-overlapping regions in space and are often referred to as marine provinces, biogeographic realms, seascapes, or ecoregions \citep{Spalding_2007,Briggs_2012,Costello_2017, kavanaugh2014hierarchical}. Coherent delineation of ocean provinces helps us better understand global-scale biogeochemical processes (e.g., primary production \citep{Reygondeau2020, Tagliabue2021}), microbial distribution and diversity \citep{Plankton-Distribution, horstmann2022microbial, Raes2018}, assess fishery biomass and global ocean colour trends \citep{palomares2020fishery, vanOostende2023}, and assist with validation of larger scale ocean biogeochemistry general circulation models \citep{Vichi2011, sonnewald2020elucidating}. There have been numerous efforts to define oceanic provinces, but much of the foundational work drew on globally distributed and seasonally varying ocean properties, such as chlorophyll concentration and parameters of the photosynthesis and light relationship throughout the water column, to partition primary production estimates into secondary compartments of 57 biogeochemical provinces nested within four primary pelagic biomes: Polar, Westerlies, Trade Winds, and Coastal \citep{Sathyendranath_1995, Longhurst_1995}. 

% \notate{YR: \citep{Westberry2013} seems irrelevant to me - it's just a chapter summarizing NPP calculations using satellite monitoring, also moved the ref. \citep{vanOostende2023} to prior sentence b/c it's actually just used to monitor trends in global ocean colour throughout different BGCPs - careful confusing satellite-based models with ecosystem modelling: also, I found a more relevant citation for that angle btw \cite{Vichi2011}}

% \notate{SH: Yubin will edit this paragraph.}
Longhurst \cite{longhurst-book} transformed the original province concept into a comprehensive biogeographical framework by systematically integrating a wide range of oceanographic parameters—including water column structure, nutrient availability, plankton community composition, and large-scale circulation patterns. Rather than relying solely on surface properties like chlorophyll or irradiance, Longhurst \cite{longhurst-book} incorporated data from hydrographic sections, climatological atlases, and ecological surveys to align province boundaries with functionally distinct oceanic regions. These refinements anchored the provinces not just in satellite observables but in ecological and physical processes, establishing a globally coherent system that reflects the dynamic coupling between ocean physics, biogeochemistry, and plankton ecology. Other studies have also aimed to objectively improve on these oceanographic boundaries by investigating their dynamic nature  \citep{Hooker_2000,Oliver_2008,Reygondeau_2013, kavanaugh2014hierarchical} and integrating higher trophic organisms \citep{Briggs_2012,Costello_2017}. More recent efforts have also sought to improve on these provinces by partitioning the deeper ocean layers such as the mesopelagic zone \citep{Sutton_2017} or by incorporating more highly resolved phytoplankton biogeography \citep{Elizondo_2021}. 

%There have been multiple efforts in defining ocean provinces. The foundational work which defined oceanic provinces by leveraging physical and chemical hydrographic parameters \citep{Longhurst_1995, Sathyendranath_1995, Hooker_2000}.  These studies defined provinces with static boundaries of surface seawater, and were developed using satellite and in-situ data such as chlorophyll concentrations, ocean color, mixed layer depth, nutrient availability, and seasonal cycles, as well as physical and oceanographic features such as ocean currents, upwelling zones, and fronts \notate{SH: The classical works didn't use all of this; we should make this clear.}. 
 
%\notate{SH: Say something like ``the literature culminated into \cite{Longhurst_2010}, which acknowledges the importance of seasonal cycles, etc."}

With the advent of more affordable molecular technology and sequencing platforms, there is a proliferation of global ocean plankton surveys which provide scientists with incredible insight into the vast taxonomic and functional diversity of \textit{in situ} marine plankton compositions \citep{Duarte_2022,Sunagawa_2020,McNichol_2021,Vincent_2022}. These datasets contain marine plankton relative abundances from the global ocean across a wide range of latitudes, longitudes, and depth, providing useful data that can be incorporated into ocean province boundary estimates. One notable related work \citep{Milke_2023} grouped prokaryotes together into `modules' using co-occurrence networks, and investigated the spatial distribution of these modules.  However, to our knowledge, no effort to date has (1) incorporated biological distribution of plankton at a molecular level with taxonomic granularity and (2) across depth in concert with abiotic environmental parameters into determining these provinces and oceanographic boundaries.

%However, there has been no effort to combine these compositional plankton survey data with abiotic variables and incorporate them into a clustering framework which would help draw oceanographic boundaries. %\notate{YR: this sentence is very redunant to the last sentence of the previous paragraph - we need to pick one of them and I believe ending this section with this message creates for a better flow.} 

In this manuscript, we have developed a novel \texttt{bioprovince} algorithm (and accompanying open-source R software), which enables estimates of three-dimensional provinces extending across latitude, longitude, and depth in the ocean, primarily based on biological compositions of organisms. This method, which directly clusters biological samples and predicts spatial provinces, uses statistical and machine learning techniques in a pipeline with the major components being \textit{objective}, in the sense that the key hyperparameters can be tuned from the data. This is highly preferable to a method that requires \textit{subjective} choices from the data analyst, and ensures that the analysis results are reproducible.  Consequently, the bioprovince framework is broadly applicable to any compositional dataset, such as those generated from metabarcoding, that is observed over space and is paired with associated abiotic metadata.

In order to prove the utility of our method, we applied it to compositional microbial phytoplankton data, which we sourced from the Global rRNA Universal Metabarcoding Plankton database (GRUMP) \citep{mcnichol_characterizing_2025}.  This dataset consists of sequenced DNA, which once bioinformatically processed, produce Amplicon Sequence Variances (ASVs). Each ASV or ``DNA tag" is unique and belongs to a plankton species. The gene abundance or counts of each ASV in the GRUMP dataset are first transformed into relative abundances, based on our view that the relative abundance reflects the community mix of microbe species in the ocean. Off-the-shelf clustering methodology is usually not suitable for simply applying to compositional data, as most clustering methods assume a normal, Euclidean geometry of the data. Also importantly, clustering ASV samples with off-the-shelf clustering methods does not guarantee that spatially coherent clusters of samples or provinces emerge. Our method incorporates spatially homogeneous clusters explicitly, and performs province predictions on a fine grid based on abiotic similarity. In our manuscript, we have made the following contributions:
\begin{enumerate}
\item We have developed an objective pipeline (\texttt{bioprovince}) for estimating ocean provinces across latitude-depth given ASV compositions and simple abiotic data (temperature and salinity).
\item We have developed methodology for tuning the main hyperparameters of our method. (Section~\ref{sec:method})
\item We have applied our method to compositional phytoplankton ASV data from five cruises that encompass north-south transects in the Pacific Ocean across 14 years.
\end{enumerate}
Our pipeline for clustering of biological data (the \texttt{biocluster} function) and province predictions on a spatial grid
(the \texttt{bioprovince} function) are freely available as an \texttt{R} package, on github
\url{https://github.com/sangwon-hyun/bioprovince-project}.
%\item We develop software and a web application \marginnote{\tiny Web app needs to be rebuilt} for easy application to any compositional dataset collected over space (e.g., latitude and depth).

%\item Difficulties: Compositions are notoriously hard to work with. List other methodological difficulties.
%\item Highlight shortcomings of the existing clustering methods when applied to this dataset.  
%\item Highlight the interdisciplinary aspect (ocean + statistics)

%\item (Make sure to mention Clustgeo \cite{RN35} here.)

\begin{figure}
\includegraphics[width=.62\linewidth]{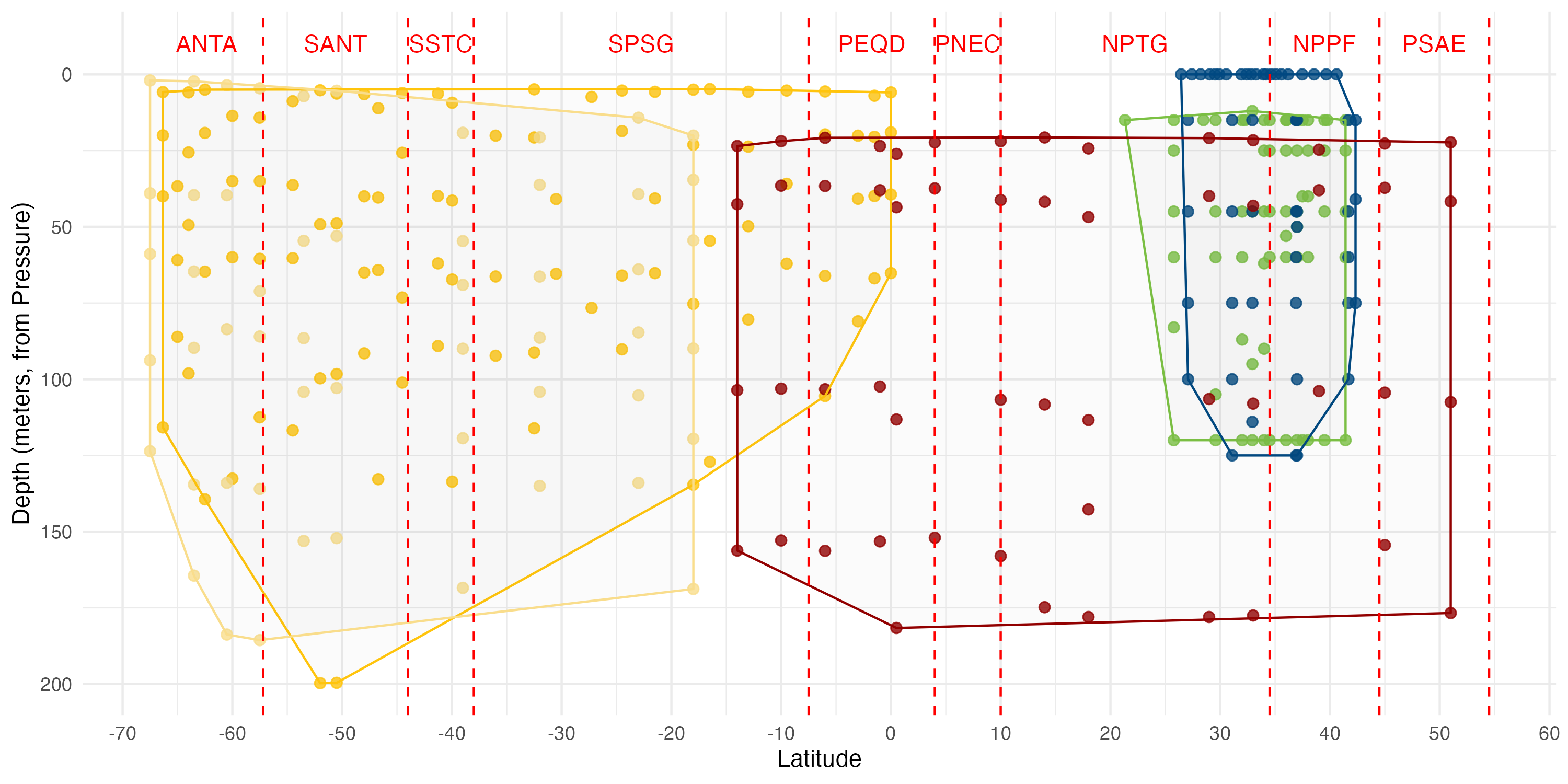}
\includegraphics[width=.37\linewidth]{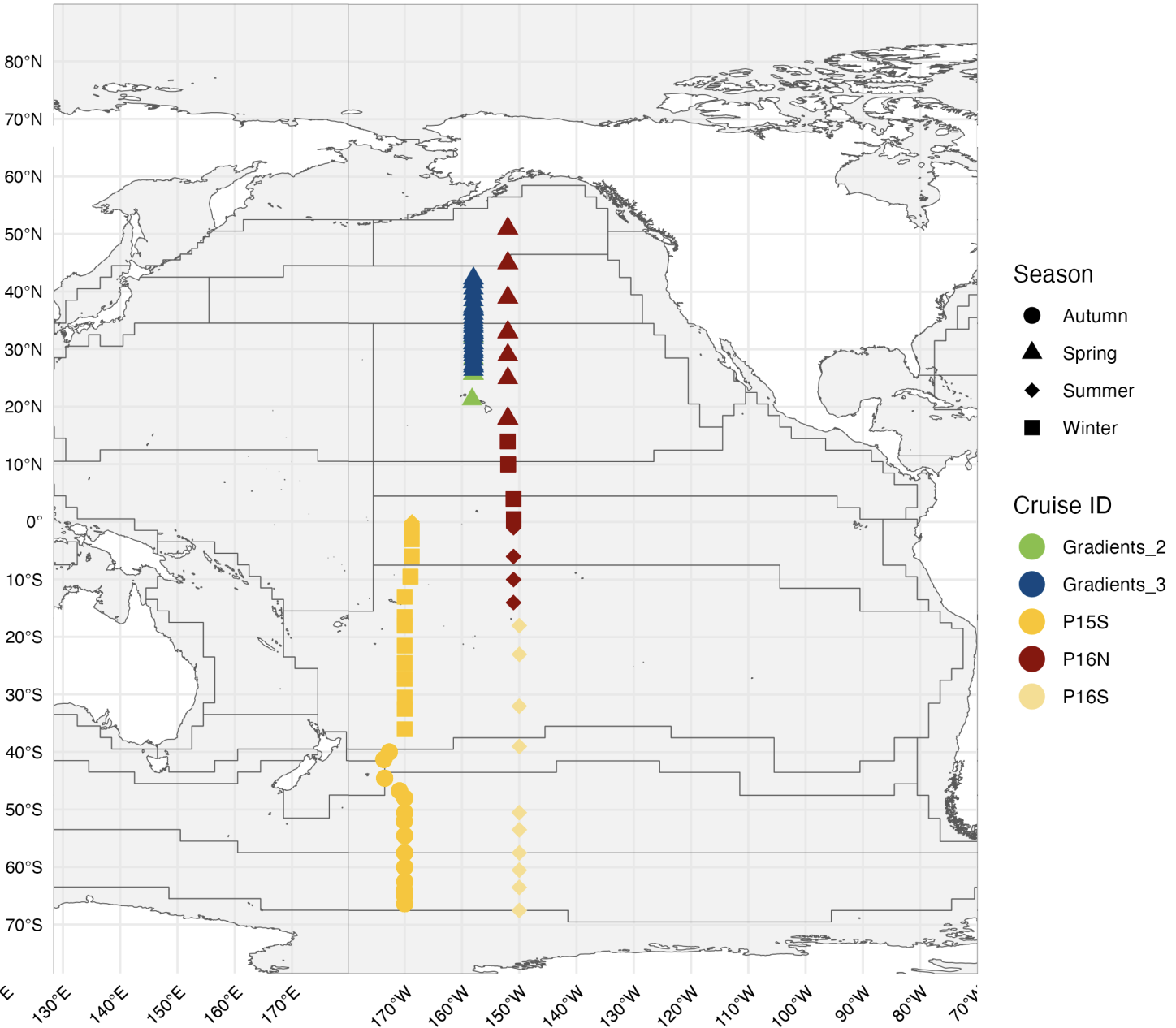}
\caption{\it Map of samples in space (latitude, longitude, and depth). There is heavy overlap between the cruises' latitude-depth coverage. The two types of cruises (P15S, P16S, P16N compared to Gradients 2 and 3) differ greatly in their sampling.  Note that we include data from 200m or above, which includes samples from below the euphotic zone.}
\end{figure}
%\notate{SH: two shades of yellow color aren't different enough. ALSO importantly, maybe we'll draw colored bars (like in the style of confidence intervals) of newly discovered provinces -- perhaps only the ones along the surface, or overlapping provinces over all depths!} \notate{SH: add date range to right-hand-side panel of Figure 1, directly in the plot using annotate().}

% -- collected from several different oceanographic cruise missions over the years -- but our method can easily be applied to a broader compositional dataset in the wider ocean. %(We focus on Phytoplankton in this paper, but the methodology generalizes to {\it any} oceonagraphic microbe composition measured over space).

\section{Material and Methods}
\label{sec:method}
%TODO: Justin will write a draft.

%In compositional data analysis, we care about relative abundance, or proportions of a whole. 

%\marginnote{\tiny
%TODO: 
%\begin{itemize}
%\item establish notation for a vector?
%\item clarify why we use an aitchison distance? Possibly remove reference to aitchison geometry, or greatly clarify this.
%\item Explain why we need to establish distances, and introduce the notation of a pairwise distance matrix.
%\item Write about hierarchical clustering?
%\end{itemize}
%}

%A curious fact is that this distance is equivalent to using Euclidian distance after applying the  Central Log-Ratio transformation to the data, defined as
%\begin{equation}
%    \textnormal{clr(x)} = \left[\log \frac{x_1}{g(x)} \ldots \log \frac{x_p}{g(x)}\right],
%\end{equation}
%where $x$ is a composition vector and $g(x)$ its geometric mean. 

%For both of our Goals, we use \ref{eq:ait_dist} for different purposes. For AIM 1 after having transformed the data, for each ASV we have a vector that represents how relatively abundant this ASV was in relation to its relative abundance in each sample. For AIM 2, we used Aitchison distance to represent how unsimilar a pair of samples are in relation to its' biotic information  $X$. 
%

\subsection{Data}

%Furthermore, these ASVs were produced using the universal 515F/926R primer pair \citep{parada2016every}, that amplifies both 16S and 18S rRNA from prokaryotes and eukaryotes, which allowed us to analyse the abundance of ASVs, between samples, and across domains of life. This is particularly important for analysis of phytoplankton -- the focus of our current study \notate{SH: Yubin will add a bit here.} -- which include prokaryotic cyanobacteria as well as larger eukaryotic phytoplankton. 

For our analysis, we used the "Global rRNA Universal Metabarcoding Plankton" (GRUMP) database version 1.3.5 \citep{mcnichol_characterizing_2025}. GRUMP consists of 1,194 samples that were collected between 2003 and 2020 and cover extensive latitudinal and longitudinal transects, as well as depth profiles in all major ocean basins. Two significant factors make these data "universal". The first is that the samples are not fractionated, meaning microbes from all size fractions are collected on the same filter during the sampling process. The second is that the DNA is sequenced using a universal primer pair, which captures all three domains of life. This means that the relative abundance data between all organisms and all samples within the dataset are comparable. 

For this study, we used samples from five different cruises that latitudinally transected the Pacific Ocean. These cruises included the P15S cruise, which occurred in 2016, the P16S cruise in 2005, the P16N cruise in 2006, the Gradients 2 cruise in 2017, and the Gradients 3 cruise in 2019 (Figure 1). These cruises span the equatorial to high-latitude Pacific and were conducted in different years and seasons, providing a rich set of spatiotemporal contexts for evaluating community structure. Two of these cruises, P15S and P16S - offset by $\sim$20 degrees longitude, traversed a vertical latitudinal transect from around the equatorial Pacific down into the highly productive waters of the Southern Ocean, but took place more than a decade apart (P16S in 2005, P15S in 2016).The other three cruises (P16N, Gradients 2, Gradients 3) transected the North Pacific Ocean at various spatial scales, spanning from the equator to $\sim$50N and were also taken multiple years apart (P16N in 2006, Gradients 2 in 2017, Gradients 3 in 2018). The two SCOPE-Gradients cruises provide much smaller spatial coverage north of Hawaii, although at a much higher sampling frequency across the North Pacific Transition Zone (NPTZ) which is a significant ecological region separating the warmer, oligotrophic gyre and colder, more nutrient rich subarctic gyre \citep{Dutkiewicz2024Multiple}. It also differs from the GO-SHIP cruises in that a bidirectional transect is conducted so that the same latitudinal region is sampled twice separated over the course of about two weeks. %Notably, seasonality also plays a major role in our analysis in this section. 
\par

Seasonal differences across cruises further contribute to variability in community structure. Gradients 2 (May and June 2017) and 3 (April 2019) were conducted two years apart during overlapping seasons in the boreal spring (and also transition to early-summer), when there is often an uptick in phytoplankton productivity across the NPTZ, but strong environmental gradients (e.g., chlorophyll-\textit{a} fronts, sea surface temperature, nutrients) shift seasonally \citep{Dutkiewicz2024Multiple}. The two P16 research expeditions, P16S and P16N, were also conducted during similar months, January - March, in consecutive years (2005 and 2006, respectively). However, due to their sampling locations spanning both hemispheres, they covered different seasonal contexts: P16S occurred during the austral summer, while P16N captured the austral summer south of the equator and a transition from  boreal late-winter to early-spring north of the equator. The remaining cruise P15S, despite overlapping with P16S/N across similar regions south of the equator, was conducted about a decade later from April - June 2016, and captures the austral transition from autumn into winter. %Collectively, these five cruises capture significant spatiotemporal variability and offer an opportunity to evaluate the utility of the bioprovince algorithm in disentangling phytoplankton community structure across diverse biomes in the Pacific Ocean. 
\par

Seawater was collected with niskin bottles at each station and then filtered through a 0.22-um filter that captures the entire microbial planktonic community, except for free-living viruses. The cells on this filter were then lysed to remove DNA from them, and then this DNA was purified for sequencing. Relative abundance information was generated by amplifying DNA with PCR using the 515FY, 926R universal rRNA gene primer pair \citep{parada2016every} that amplify 16S (prokaryotic and chloroplast) and 18S (eukaryotic) ribosomal RNA. For this study, we used relative abundance values in the GRUMP column 'Corrected Sequence Counts'. These counts are corrected based off an essential downstream  bioinformatic correction, as well as a sequencing biased against the longer 18S sequences compared to the shorter 16S sequences. For full details on sample processing, amplification, sequencing and bioinformatic pipeline, please see \citep{McNichol_2021}. 
To assess the applicability of this statistical framework, we wanted to work with a digestible subset of the GRUMP data, which includes the three major plankton realms (i.e., bacterioplankton, phytoplankton, zooplankton). Phytoplankton are the major primary producers in the ocean ecosystem and have a profound impact on global biogeochemical cycles, supplying roughly half the oxygen in our atmosphere \citep{falkowski1998biogeochemical,field1998primary}. Additionally, their roles and contributions in the ocean ecosystem are a primary component of classical Longhurst provinces and therefore, to stay consistent with our application of this algorithm, we also focused our efforts on the phytoplankton community in GRUMP. 

%The phytoplankton were filtered out of the dataset by making a subset of the GRUMP data which first included the following Cyanobacteria; Prochlorococcus, Synechococcus, Richeria, Trichodesmium, UCYN-A,  UCYN-C, and Crocosphaera. Also included in this subset were all chloroplast 16S sequences excluding Rhizaria, Excavata, Alveolata, Rhodophyta, and Dinoflagelletes. At this stage, we acknowledge that many Dinoflagelletes are considered phytoplankton, but at this stage we are unable to tease apart the phytoplankton from the non-phytoplankton, and so they were excluded from this data analysis. 

%\notate{YR: reframe how the phytoplankton community was defined from the dataset. 
%\notate{SH: Yubin will mention why we chose phyotplankton.}
This phytoplankton community was defined from ASV data as a combination of select Cyanobacteria and eukaryotic taxa inferred from chloroplast 16S sequences. The cyanobacterial component includes some of the smallest and most abundant photosynthetic organisms, such as \textit{Prochlorococcus} and \textit{Synechococcus}, as well as common nitrogen-fixing taxa (diazotrophs) such as (e.g., \textit{Trichodesmium, Crocosphaera}). Eukaryotic phytoplankton were identified based on chloroplast 16S sequences, indicative of their photosynthetic potential, with several major lineages excluded from the analysis. Specifically, sequences affiliated with macroalgae (e.g., Rhodophyta-red seaweeds), groups that commonly include kleptoplasts and photosymbionts (e.g., Rhizaria, Excavata), and Alveolata (which broadly include mixotrophic lineages in addition to kleptoplasty and photosymibioses) \citep{Dorrell2012}. Notably, even though many dinoflagellates can be important constituents of the phytoplankton community, this group has lost most of the chloroplast genes \citep{LilaKoumandou2004} and therefore, is not very well captured in this sequence fraction of the dataset. Also, while dinoflagellates are well represented in the nuclear (18S) fraction of the GRUMP database, we are unable to tease apart their trophic strategy, which includes heterotrophy and mixotrophy, so they were excluded from this data analysis.

\subsection{\texttt{bioprovince} Algorithm}
\label{subsec:algorithm}
Our proposed method, which we coin {\tt bioprovince} algorithm, first creates geographically coherent clusters of compositional samples, then defines provinces in latitude and depth according to their agreement with the original samples' cluster memberships as well as their abiotic factors. Our method proceeds in three sequential steps:

\begin{enumerate}
\item[{\bf Step 1.}] Form a pairwise distance matrix of samples. Each distance
  is a convex combination of (1) biological distance between the two compositions, and (2) spatial distances between the sample's location.
\item[\bf{ Step 2.}] Use agglomerative hierarchical clustering to assign cluster memberships to the
  samples. We call Step 1 \& 2 the {\tt biocluster} algorithm.
\item[\bf{ Step 3.}] On a fine set of spatial gridpoints, use a {\it localized} nearest neighbor
  classification using abiotic factors.
  %\notate{SH: I'm considering adding as Step 4 the subsampling approach.}
\end{enumerate}
\noindent Data analysts who wish to cluster samples can apply \texttt{biocluster} in Step 1-2.
Those who wish to further constrain the spatial grid to define novel provinces should proceed to Step 3. We describe each step in detail next.
\paragraph{Step 1: Form pairwise distance matrix.} 
Denote the compositional dataset as $\{\mathbf p_i\}_{i=1,\cdots,n}$, where $\mathbf p_i$ is a vector containing $d$ microbial compositions observed at the $i$-th sample. Form pairwise biological distance matrix 
$D_{\text{bio}} \in \R^{n \times n}$ of the $n$ samples using an {\it
  Aitchinson} distance \cite{aitchison1982} of the $n$ compositions $\mathbf p_i$ (for $i=1,\cdots,n$), so that the
$(i_1,i_2)$-th entry $D_{\text{bio}}[i_1,i_2]$ is:
\begin{equation}
\label{eq:aitchison-mat}
D_{\text{bio}}[i_1,i_2] = d_{\text{Aitchison}}(\mathbf p_{i_1}, \mathbf p_{i_2}).
\end{equation}
The Aitchison distance between two $d$-dimensional compositional vectors $\mathbf p = (\mathbf p_{[l]})_{l=1,\cdots, d}$ and $\mathbf q=(\mathbf q_{[l]})_{l=1,\cdots,d}$: 
\begin{equation}\label{eq:ait_dist}
  d_{\text{Aitchison}}(\mathbf p, \mathbf q) = \sqrt{\sum_{l=1}^{d} \left(\ln\frac{\mathbf p_{[l]}}{g(\mathbf p)} - \ln\frac{\mathbf q_{[l]}}{g(\mathbf q)}\right)^2},
\end{equation}
where $g(\mathbf x) = \sqrt[d]{\prod^d_{l=1} \mathbf x_{[l]}}$ of vector $\mathbf x=(\mathbf x_l)_{l=1,\cdots, d}$ is the geometric mean of its entries.

Also, denoting as $\lat(\text{sample\;}i)$ and $\depth(\text{sample\;}i)$ the latitude and depth of sample $i$, form the spatial distance matrix:
\begin{equation}
\label{eq:spatial-mat}
D_{\text{spatial}}[i_1,i_2] = \sqrt{
r^2 \cdot
(\lat(\text{sample\;}i_1) - \lat(\text{sample\;}i_2))^2 +  (\depth(\text{sample\;}i_1) - \depth(\text{sample\;}i_2))^2}.
\end{equation}
Next, take a convex combination of $D_{\text{bio}}$ and $D_{\text{spatial}}$ as
follows (denoting as $h()$ the scaling operation that divides the columns of
a matrix by its operator norm):
\begin{equation}
\label{eq:mix-matrix}
D_\alpha = (1-\alpha) \cdot h(D_{\text{bio}}) + \alpha \cdot h(D_{\text{spatial}}).
\end{equation}
for some fixed hyperparameter $0\leq \alpha \leq 1$.

We note that an existing method ClustGeo \citep{RN35} takes a similar approach
to clustering spatial data. However, step 1 of our method is specialized for
the application of marine microbe compositions, since the biological distances
$D_{\text{bio}}$ are Aitchison distances of compositional data, and the spatial
distances -- in latitude and depth -- require additional care to determine the relative importance ($r$) of the two. We suggest a novel data-driven tuning strategy for the various hyperparameters ($\alpha$, $r$, and $K$), described shortly in Section~\ref{subsec:tune}.

\paragraph{Step 2: Cluster the samples.} Using the pairwise distance matrix $D$ from \eqref{eq:mix-matrix}, perform agglomerative hierarchical clustering \citep{Sokal_1958} to obtain cluster memberships for each of the $n$ compositions. Agglomerative hierarchical
clustering is a bottom-up approach, starting with $n$ distinct clusters,
combining the two that have the smallest pairwise distances, then successively
merging {\it hierarchically} using a linkage function between {\it groups} until
$K$ groups emerge. We opt for the Ward linkage function \citep{Ward1963}. 
%The resulting estimated cluster membership of sample $i$ is 
%\marginnote{\tiny Do we need this notation?}
%$$z_i \in \{1,\cdots, K\},$$
%for sample $i=1,\cdots,n$. (Note, the number of clusters $K$ can be tuned, as is explained in the
%next section)

%take as $I_1 \cdots I_n$ to be the indices of the sorted spatial distances $d_1,\cdots, d_n$, from small to large. A nearest-neighbor prediction is then made by 
\paragraph{Step 3: Localized nearest neighbor prediction of memberships on a grid.} 
Take a $B$-sized fine grid of spatial locations, indexed by $j$ in $1$ through $B$.  For each of these spatial grid points $j$, find the $k$ most similar samples according to {\it abiotic} conditions. Specifically, take the abiotic similarity between grid point $j$ and sample $i$ to be:
%$$d(j, i) = d_{\text{abiotic}}(j, i),$$
$$\sqrt{(\text{temperature}(\text{grid\;} j) - \text{temperature}(\text{sample\;}i))^2 + (\text{salinity}(\text{grid\;} j) - \text{salinity}(\text{sample\;}i))^2},$$
the $\ell_2$ difference between temperature and salinity measurements  -- both linearly rescaled to be between zero and one. Next, take the estimated cluster memberships of these $k$ closest samples that are \textit{also} within the $k$ closest in space, where spatial distance is defined as before in \eqref{eq:spatial-mat}:
$$ \sqrt{r^2 \cdot (\lat(\text{grid\;}j) - \lat(\text{sample\;}i))^2 + (\depth(\text{grid\;}j) - \depth(\text{sample\;}i))^2}.$$
If there are no such samples, take the single spatially nearest sample's membership. Call these memberships (each a number out of $1,\cdots, K$):
$$ z_1,  z_2, \cdots,  z_{k'}.$$
where $k'$ can be as low as $1$, and as high as $k$. Then, predict the membership $\hat z$ of the $j$th grid point by a majority vote of these memberships:
$$ \hat z({\text{grid}\;}j) = \text{mode}( z_{1}, \cdots,  z_{k'}),$$
and when there is a tie, choose the single nearest neighbor's membership. Repeating this for each grid point $j=1,\cdots, B$, this process determines predicted province memberships
$\hat z (\text{grid\;} 1),\cdots, \hat z(\text{grid\;}B)$
for every grid point $j=1,\cdots, B$.  
(An example is shown in Step 2 and 3 of Figure~\ref{fig:workflow}.)

\subsection{Assessing stability of newly learned provinces}
\label{subsec:stability}

Once an analyst has newly learned provinces on a given dataset, it is useful to quantify the associated uncertainty. In order to assess how {\it stable} an estimated province is, we devise a subsampling strategy for the dataset. Specifically, we randomly subsample $70\%$ of the ASVs in a dataset, and calculate how often the predicted memberships on a grid persist across subsampling. This method proceeds in these four steps:
\paragraph{Step 1.} For each compositional sample $\mathbf p_i$ (for $i=1,\cdots, n$), resample $70\%$ of the ASVs in $\mathbf p_i$ and renormalize to sum to one, to obtain a shorter $\mathbf p_i^{(l)}$. Repeating this $100$ times (for replicates $l=1,\cdots,100$) to obtain 100 subsampled compositions $\{ \mathbf p_i^{(l)}\}_{i=1,\cdots, n, l=1,\cdots 100}$.
\paragraph{Step 2.} For each new subsampled dataset $\{\mathbf p_i^{(l)}\}_{i=1,\cdots, n}$, rerun {\tt bioprovince} to obtain new province definitions at fine grid points $j=1,\cdots, B$.
\paragraph{Step 3.} Re-align the $K$ cluster labels in each of the results (so that a given cluster number $k$ is consistent across all subsampled datasets), according to their geographical similarity. 
\paragraph{Step 4.} Then, for each grid point $j=1,\cdots, B$, assess the {\it stability} of the province as the frequency with which grid point $j$ was deemed the most common province $k_j$, defined shortly.  
\\

We further describe some details. In order to realign the $K$ clusters, we apply the Hungarian matching algorithm to the pairwise differences between the $K$ clusters, where the pairwise difference between province $k_1 \neq k_2$ is measured as the average spatial distance between all samples in cluster $k_1$ and all samples in cluster $k_2$. To measure the stability, we measure the relative frequency that grid point $j$ is deemed cluster $k_j$, where $k_j$ is the most frequent cluster:
%(whose spatial locations are $s_j, j=1,\cdots B$) 
$$ k_j = \text{argmax}_{k=1,\cdots, K} \left(\#\text{times gridpoint } j \text{ is deemed cluster } k\right),$$
and
$$ \text{Stab}_j = \frac{\# \text{times gridpoint } j \text{ is deemed cluster } k_j}{100}.$$

Appendix Figure~\ref{fig:stability-p16} and \ref{fig:stability-gradients} shows an example of this, where grid points over latitude and depth are colored by the province membership $k_j$, and the opacity is proportional to $\text{Stab}_j$.

\subsection{Choosing hyperparameters}
\label{subsec:tune}

In this section, we provide data-driven strategies for determining the
hyperparameters for {\tt bioprovince}. We recommend picking them in a particular order ($r \rightarrow \alpha \rightarrow K$) prior to applying the bioprovince algorithm. 
\begin{enumerate}
\item Calculate a value $r$ (relative importance of latitude versus depth) 
\item Pick a value of $\alpha$ for clustering, using a visualization technique (described below).
\item For this $\alpha$, pick number of clusters $K$, using within-sample
  variance.
\end{enumerate}

The hyperparameter $r$ governs the relative importance of latitude (in degrees) and depth (in meters) in our proposed pipeline. Since each cruises' data covers widely different spatial ranges, the use of a single $r$ parameter harmonizes the spatial scale of clustering among cruises. We devise a method resemblant of distance decay \cite{Nekola1999} from macroecology, which proceeds as follows. First take all pairs of samples' (a) biological distances, (b) latitude difference, and (c) depth difference.
Next, take only the sample pairs that are reasonably close to each other. Then, for sample pairs in each cruise, apply two separate linear regressions of (a) onto (b) and (c), and take the {\it weighted} average of the significant positive estimated slopes from each cruise -- using the number of sample pairs as relative weights. The ratio of those averaged slopes determines $r$. The full procedure is described in Appendix section~\ref{sec:supp-choose-r}.

Next, in order to quantify $\alpha$ -- the relative amount of biological and geographical information being used by clustering (in Step 1) -- we develop a visualization strategy that allows the user to pick an appropriate value. First create four additional synthetic samples, whose spatial locations are at the four corners of the latitude-depth rectangle of the existing samples, and whose biological information are randomly drawn from samples in the existing dataset. Having compressed this augmented dataset into two dimensions using multidimensional scaling (MDS), take the proportion of points (out of $n$) that are contained within a polygon whose vertices are those four points. This is a randomized metric (between 0 and 1) that measures how {\it saturated} with spatial information the distance matrix $D$ from \eqref{eq:mix-matrix} has become. We also form a ``null'' version of this metric (described in Appendix~\ref{sec:supp-choose-alpha}. In short, non-null scores use the four corners in space as an actual reference, while the null scores pick four random spatial points.  Both the null and non-null metrics can be recalculated multiple times, where the replicates come from repeatedly picking four random values (from the data itself) of biological compositions for those four corners. The value $\alpha$ (between 0 and 1) should then be chosen to be a value where the null and non-null values are still highly overlapping.

Lastly, we describe how to determine a suitable number of clusters $K$. For each candidate value of $K=1,2,\cdots$, conduct clustering (using \texttt{biocluster}), then calculate a within-cluster distance according to the mixture distance \eqref{eq:mix-matrix}:
$$\text{Average within-cluster distance} = \sum_{k=1}^K\sum_{i_1 \neq  i_2,\; i_1,i_2\in C_k} D_\alpha[i_1, i_2], $$
for $D_\alpha$ defined in  \eqref{eq:mix-matrix}. Then, choose a suitable value of $K$ visually, by identifying an ``elbow'' location where the improvement in the average within-cluster distance sharply diminishes as $k$ increases. The first panel of Figure~\ref{fig:workflow} shows an example applied on data from the P16S cruise.

% once these are done, (? is this how we use it?)
Lastly, after the \texttt{biocluster} sample clustering is done with these chosen hyperparameters, we recommend choosing the value of $k$ -- for $k$-nearest-neighbor prediction --  using visual inspection of the resulting province estimates. A small integer $k$ between 1 and 5 is usually suitable and creates spatially coherent provinces; a larger $k$ is useful when the estimated cluster memberships at the samples substantially overlap in space, and all clusters consist of a sufficiently large number of samples.

%\subsection{Software}
%
%\begin{itemize}
%\item Dashboard: Publicly available web-based software application for {\tt bioprovince}.
%\item R package: {\tt bioprovince} {\tt R} package, built using literate programming {\tt litr}.
%\end{itemize}

\section{Results}

\subsection{Application to one cruise} 
\label{sec:one-cruise-application}

\begin{figure}[htp!]
\includegraphics[width=\linewidth]{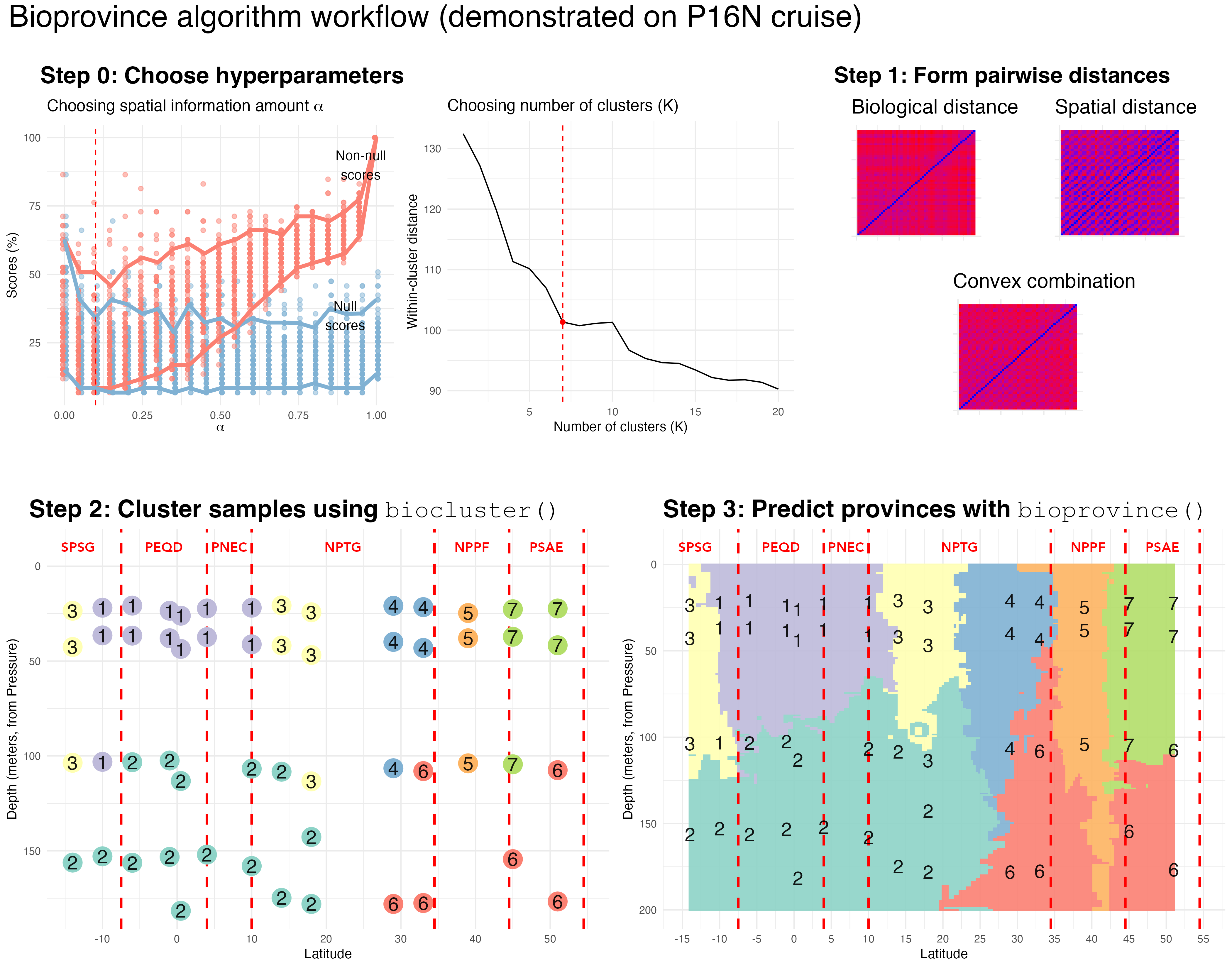}  
\caption{\it Applying the pipeline to a single cruise (P16N), as described in Section~\ref{subsec:algorithm}. The first panel describes the hyperparameter tuning described in Section~\ref{subsec:tune}. The second panel shows an example of forming a convex combination of distances between samples. The third panel shows the estimated clusters at the sample locations, shown as numbered points, obtained using the \texttt{biocluster} algorithm. The fourth, last panel shows the result of applying the \texttt{bioprovince} algorithm to conduct a localized nearest-neighbor 3D-province prediction on a fine spatial grid. The 3d-province predictions are shown in discrete colors.}
\label{fig:workflow}
\end{figure}

We first applied our pipeline to a single cruise, P16N, from the GRUMP dataset. The three panels in Figure~\ref{fig:workflow} show the result of applying each step of our \texttt{bioprovince} algorithm. Step 0 (top-left panel) shows the simulated scores (Appendix~\ref{subsec:tune}) -- both null and non-null). We pick a value of $\alpha=0.1$ for which there is substantial overlap between these two scores' distributions. Then, in the second part of Step 0 we use the within-cluster distance (Section~\ref{subsec:tune}) across candidate values of $K=1,2,\cdots$, and in this case, we identify an elbow at $K=7$. 
\par

Next, we determine how different pairs of samples should be by forming a pairwise distance matrix that is, entrywise, a convex combination of biological distance and spatial distance (using an $\alpha$ and $1-\alpha$ mix). Then, we apply our clustering algorithm \texttt{biocluster} with these hyperparameters to estimate cluster memberships of the samples as shown in Step 2 (left bottom panel of Figure~\ref{fig:workflow}). Lastly, in Step 3, we perform \texttt{bioprovince} -- a nearest-neighbor prediction of memberships -- on a fine grid in latitude-depth and based on localized abiotic similarity, to define seven different biological provinces. 
\par

The resulting provinces capture strong spatial and vertical structure in the Pacific. Two well-defined oligotrophic clusters occupy subtropical surface waters, while higher-latitude and deeper waters are divided into several additional provinces. Each province forms a coherent, internally consistent community distinct from its neighbors, demonstrating that the algorithm delineates biologically meaningful domains along the transect without relying on predefined geographic boundaries or environmental thresholds. Having demonstrated the workflow on a single cruise (P16N), we next applied the \texttt{bioprovince} algorithm to a broader dataset of five Pacific Ocean cruises to examine biological structure across larger spatial and temporal scales.
\par

\subsection{Applying bioprovince algorithm across broad latitudinal transects in the Pacific from poles to the tropics}
\label{sec:main-application}

\begin{figure}[ht!]
\centering
\includegraphics[width = \linewidth]{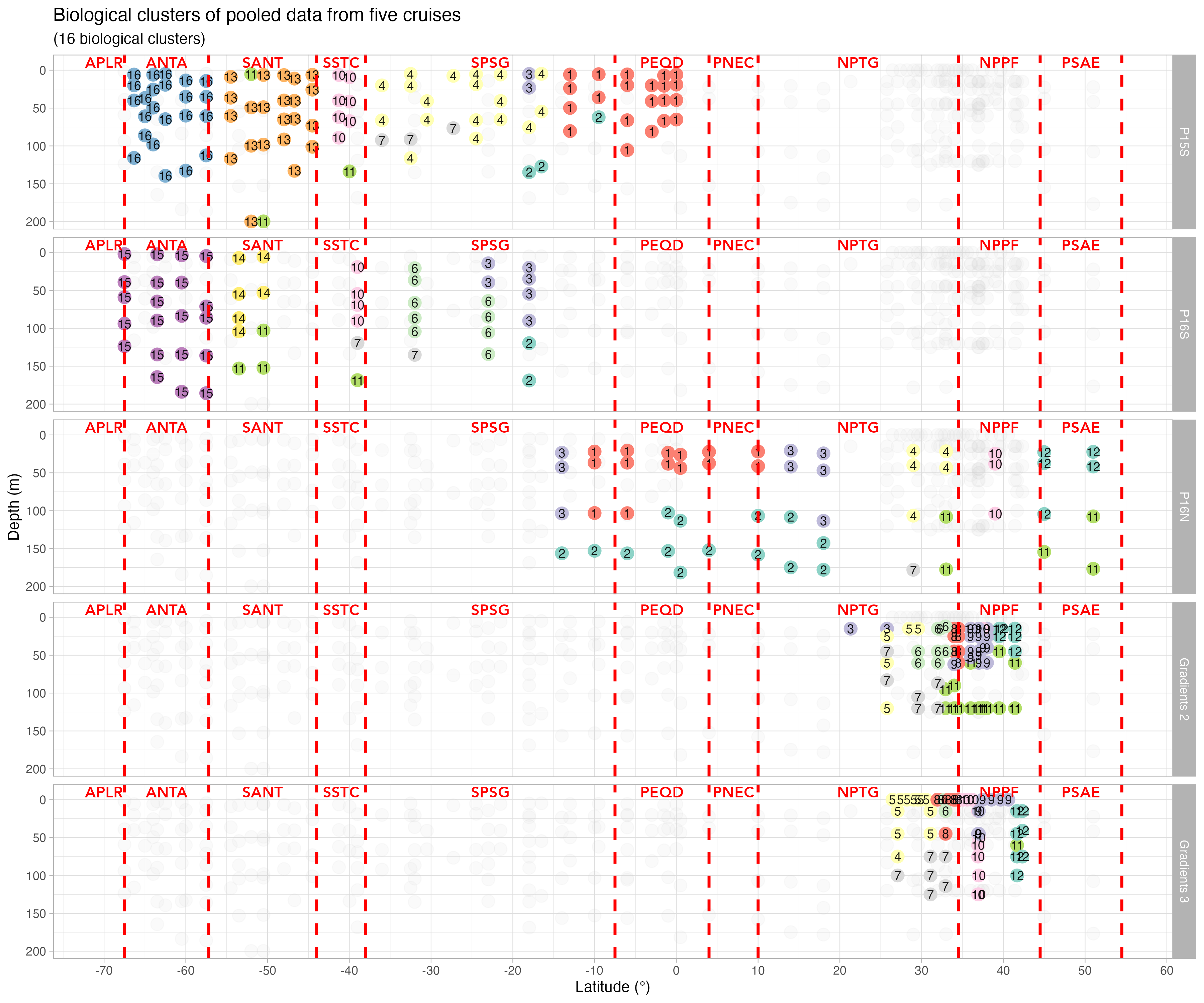}
    \caption{\it Biological clusters of samples from the pooled dataset of five cruises (P15S, P16S, P16N, and Gradients 2 \& 3), where each row shows the samples from one cruise at a time. 
    The colors and numbers show the cluster memberships estimated at the sample location, and Longhurst boundaries are shown in vertical red lines.}
\label{fig:pooled}
\end{figure}

\begin{figure}
\centering
\includegraphics[width = \linewidth]{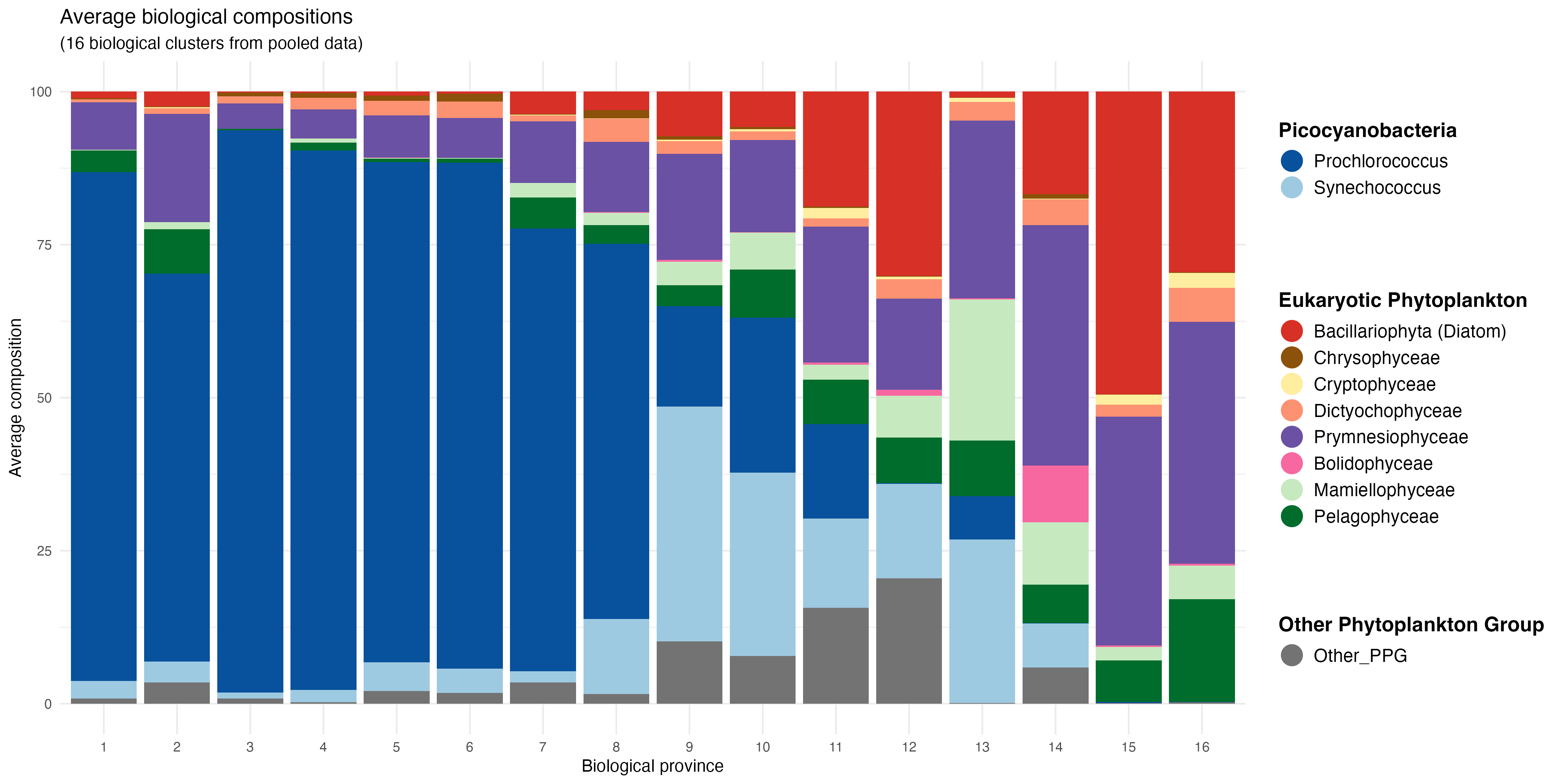}
\caption{\it (Left panel) The average relative composition for each of the 16
  biological clusters from the pooled samples from five cruises, aggregated at
  the ecologically relevant plankton group level, and visualized as
  barplots. The clusters have been ordered in ascending order of the average
  distance away from the equator.}
\label{fig:barplot-pooled}
\end{figure}

\begin{figure}
\centering
\includegraphics[width = \linewidth]{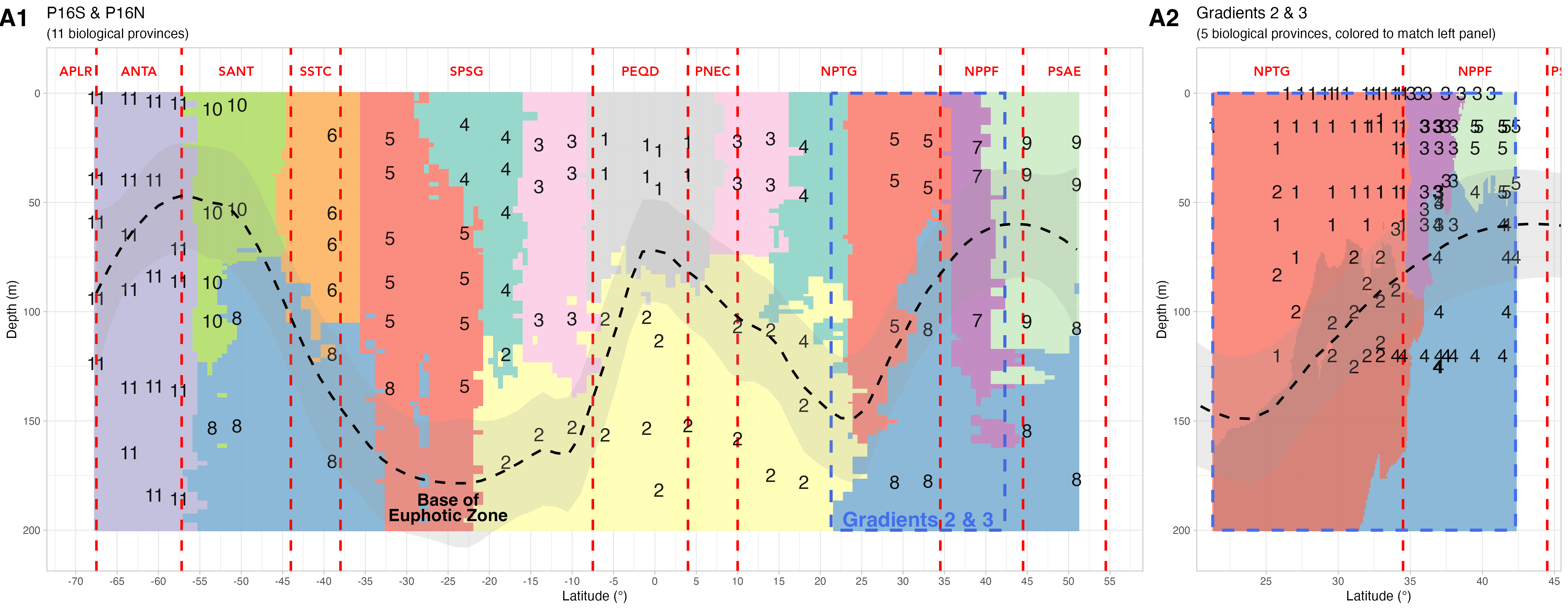}
\includegraphics[width = \linewidth, trim={0 2.5cm 0 0},clip]{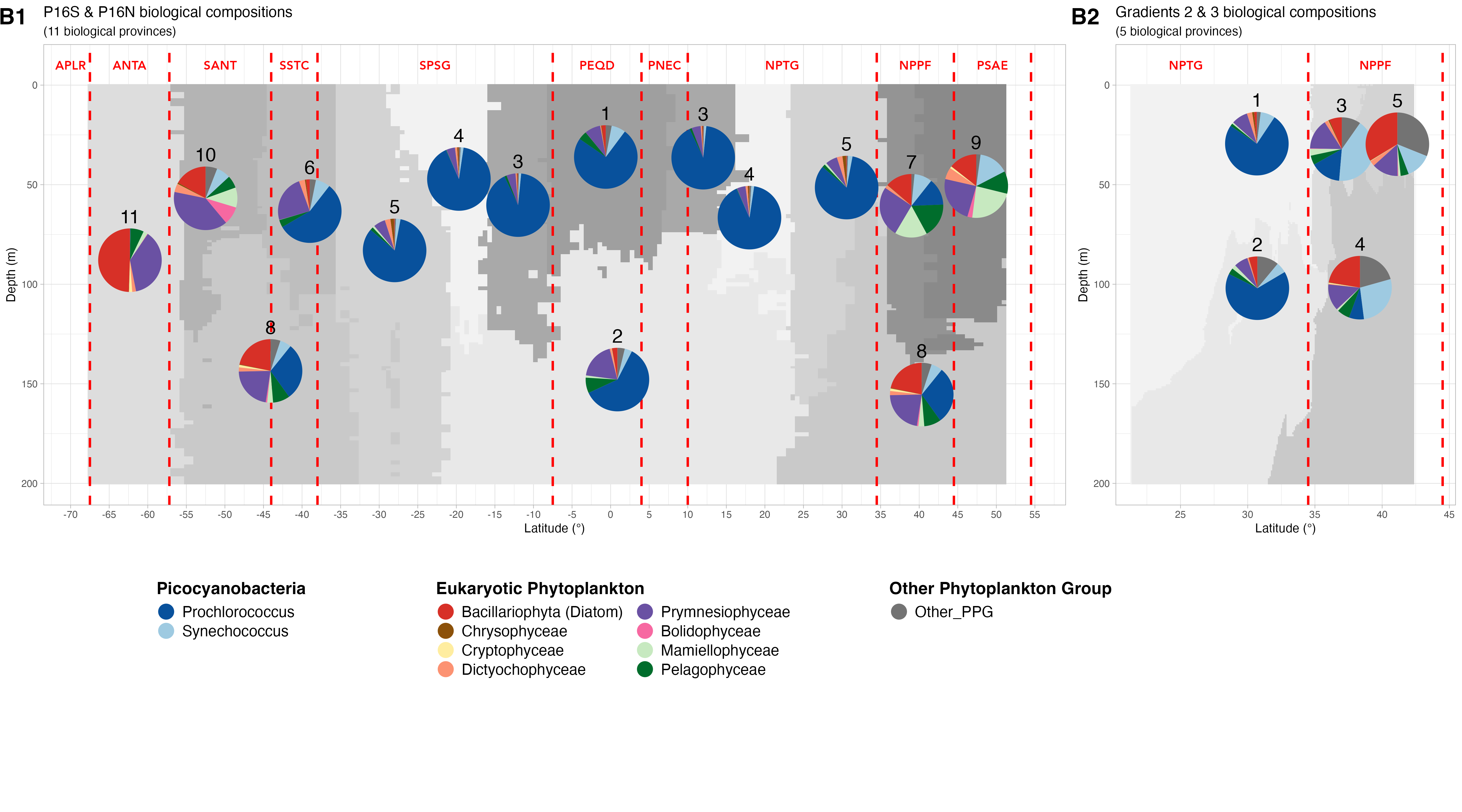}
\caption{\it  Comparing the estimated provinces from applying bioprovince to two
  cruises at a time (P15S \& P16N, and Gradients 2 \& 3). [Top row] Colors show
  the estimated province, and the numbers show the cluster memberships estimated
  at the sample locations. Longhurst boundaries are shown in vertical red lines,
  and three oceanographic fronts are shown in grey boxes. Also, the dashed black
  line shows the estimated base of the Euphotic Zone, calculated from the Darwin
  simulation model as the depth at which the sunlight (PAR) is 1\% of the
  surface level. The shaded black ribbon region shows the standard error for
  this dashed line.  [Bottom row] The average relative composition for each
  bioprovince, at the ecologically relevant plankton group level, is shown
  overlaid over greyscale background in which each province is now shown as a
  greyscale shade. Barplots of average compositions can be found in Appendix
  Figure~\ref{fig:novel-provinces-from-pairs-of-cruises-barplots}.}
\label{fig:novel-provinces-from-pairs-of-cruises}
\end{figure}

\begin{figure}
\centering
\includegraphics[width = .8\linewidth]{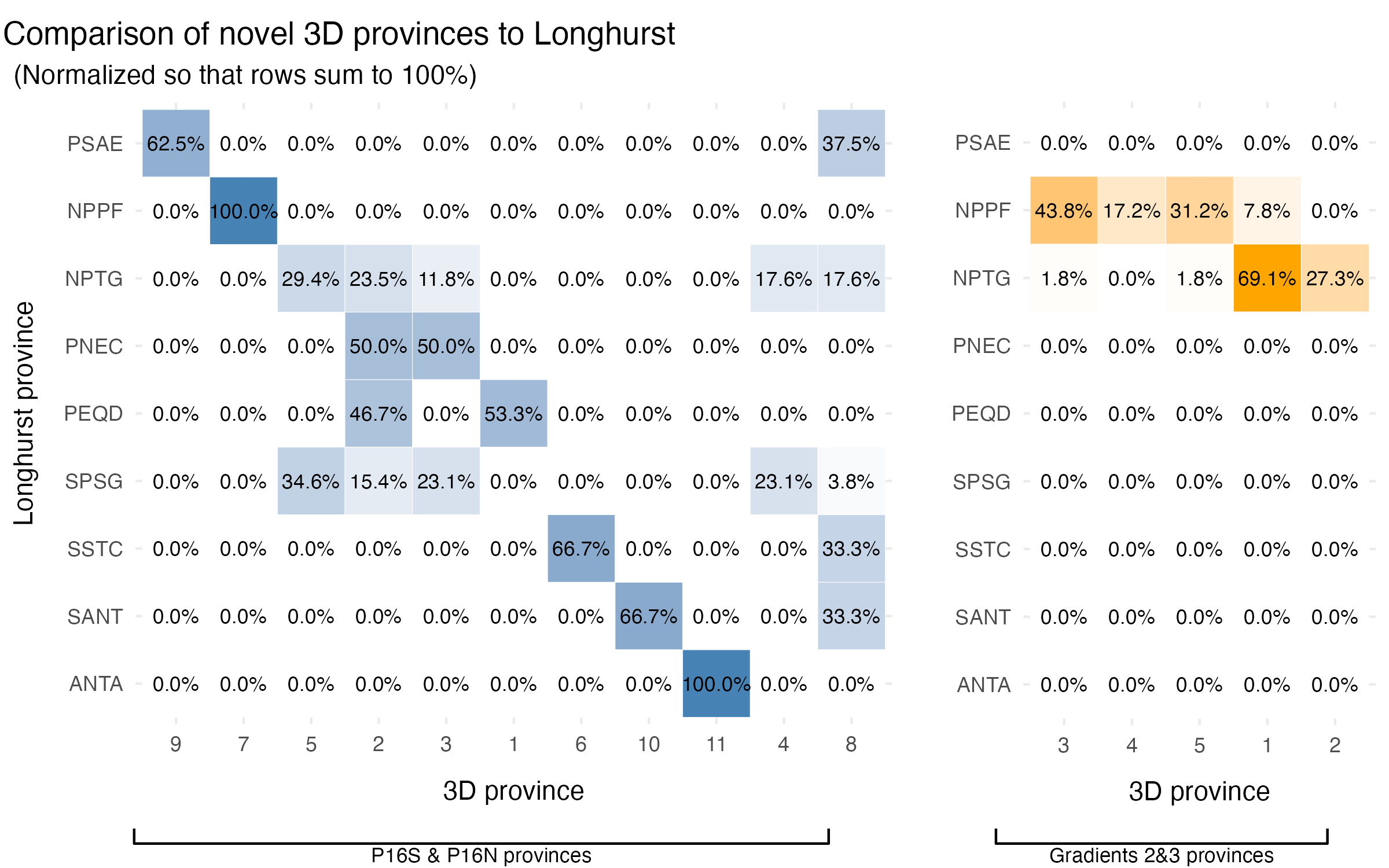}
\caption{\it Comparison of novel biological provinces to Longhurst provinces, in
  a normalized cross-tabulated table. Colors in each cell show the
  percentage. Rows (Longhurst provinces) are arranged from north to south, and
  each row's entries have been normalized to 1. The provinces have also been
  arranged from left to right according to the most highly corresponding
  province to each Longhurst province. The left panel shows the biological
  provinces from the pooled P16S \& P16N data, and the right panel shows the
  biological provinces estimated from the pooled Gradients cruise data.} 
\label{fig:heatmap-compare-to-longhurst}
\end{figure}

\begin{figure}
\centering
\includegraphics[width = .9\linewidth]{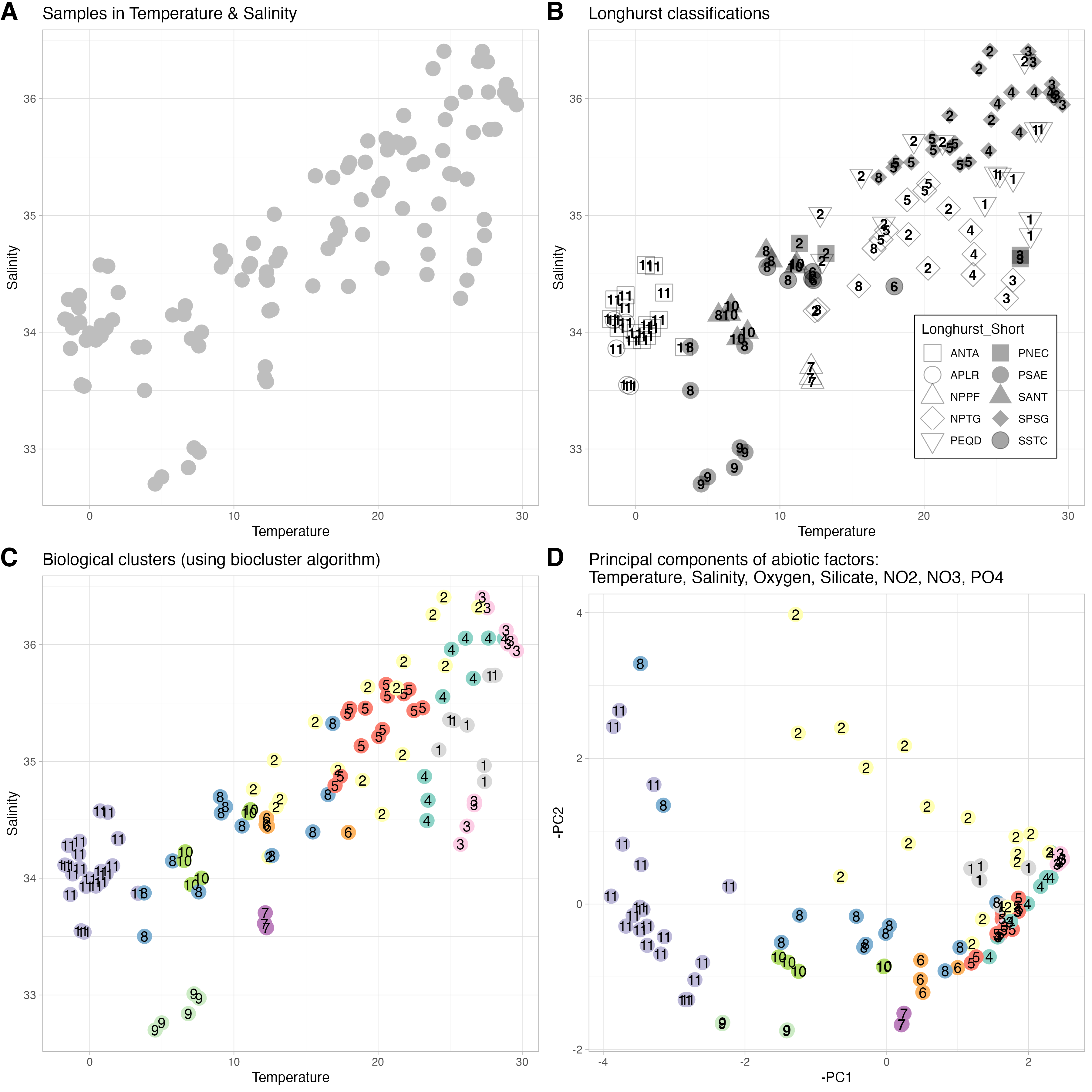}
\caption{\it (Left panel) The 11 clusters from P16S \& P16N samples in
  temperature-salinity space; the points are slightly jittered to show all
  overlapping points. (Right panel) The same 11 clusters of samples in the
  Principal Components space of five abiotic factors (Temperature, Salinity,
  Oxygen, Silicate, NO2, NO3, PO4).}
\label{fig:environment-isnt-enough}
\end{figure}

We structured the analysis in two main steps:
\begin{enumerate}
\item Pooled clustering across five cruises - We combined phytoplankton samples from all five cruises and applied the clustering algorithm to identify coherent communities across the full dataset. (See Section~\ref{sec:main-application-pooled}, Figure~\ref{fig:pooled})
\item Defined 3-dimensional provinces (which we will call 3D-provinces) from the poles to the tropics - Using pairs of cruises along latitudinal transects, we estimated biological provinces to reveal how communities vary from the equatorial Pacific to high-latitude waters. (See Section~\ref{sec:main-application-two-at-a-time}, Figures~\ref{fig:novel-provinces-from-pairs-of-cruises}, \ref{fig:heatmap-compare-to-longhurst}, \ref{fig:environment-isnt-enough}.)
\end{enumerate}

%In our main data analysis, we applied our pipeline with data-driven hyperparameters to determine biological clusters and estimate biological provinces from  phytoplankton data collected from multiple cruises within the Pacific Ocean. 

%We focus on five cruises, and conduct the following analysis:
%\begin{enumerate}
%\item We biologically cluster to five cruises' data, pooled. (Section~\ref{sec:main-application-pooled}, Figure~\ref{fig:pooled})
%\item We define novel biological provinces along north-south transect from a subset of these cruises' data. (Section~\ref{sec:main-application-two-at-a-time}, Figure~\ref{fig:novel-provinces-from-pairs-of-cruises})
%\item We compare our novel provinces principled comparisons to existing Longhurst provinces (Section~\ref{sec:main-application-longhurst-comparison}, Figure~\ref{fig:heatmap-compare-to-longhurst}).
%\end{enumerate}

%\notate{SH: refer to a figure like: Figure~\ref{fig:all-cruises}.}
\subsubsection{Biologically clustering five cruises' pooled data}
\label{sec:main-application-pooled}

Here, we applied the \texttt{biocluster} algorithm to the pooled dataset from all five cruises, resulting in 16 distinct biological clusters or ``biocluster" that are depicted within each cruise's panel (Figure~\ref{fig:pooled}). Hereafter, we adopt the notation of [cruisename]-[cluster] to refer to a biocluster in a particular cruise's panel, for notational brevity.
\par

Our algorithm found that $10$ out of the $16$ bioclusters are symmetrically distributed across the equator, confirming the intuition that the biological compositions are likely to be similar in the same environmental conditions but mirrored at the same latitudinal locations across the equator. We found the biological clustering to be \textit{robust} in the pooled dataset of multiple cruises. That is, even though the five cruises spanned highly variable dates, seasons, and spatial locations, nearby samples consistently clustered together and occupied similar spatial locations within the Longhurst provinces. (For a numerical comparison, see Appendix Table~\ref{tab:pooled-mix}). For instance, the upper $\sim$100m of the oligotrophic gyres in either hemisphere (i.e., SPSG and NPTG) are predominantly associated with bioclusters 3 (e.g., P16S-03, P16N-03), 4 (e.g., P15S-04, P16N-04), 5 (e.g., G2-05, G3-05), and 6 (e.g., P16S-06, G2-06). At a coarse taxonomic resolution, the similarity among bioclusters 3 through 6 is primarily driven by the dominance of \textit{Prochlorococcus} ($\sim$82 - 92 \%) along with co-occurrence of other picophytoplankton (e.g., \textit{Synechococcus}, pelagophytes) and nanophytoplankton (e.g., prymnesiophytes, dictyochophytes) at much lower abundance (Figure~\ref{fig:barplot-pooled}). 
\par

Also, these oligotrophic gyres are separated by biocluster 1 in equatorial waters, indicating a persistent phytoplankton community within the upper $\sim$100m (e.g., P15S-01, P16N-01). While these communities share similar picocyanobacterial members with those in the oligotrophic gyres (i.e., \textit{Prochlorococcus}: $\sim$83 \%), biocluster 1 also has an increased relative contribution from diatoms and pelagophytes (Figure~\ref{fig:barplot-pooled}). In deeper waters, we find another coherent biocluster 2 across multiple cruises (e.g., P15S-02, P16S-02, P16N-02) which is characterized by higher relative amounts of eukaryotic phytoplankton ($\sim$32 \% with a standard deviation of $\sim$14 \%) including diatoms, prymnesiophytes, and pelagophytes (Figure~\ref{fig:barplot-pooled}). Additionally, it seems that frontal zones (e.g., S. Subtropical Convergence Province – SSTC, NPPF) -- which separate oligotrophic gyres from subpolar regions -- consist of the same biocluster (P15S-10, P16S-10, P16N-10) across three cruises. The phytoplankton composition of biocluster 10 (Figure~\ref{fig:barplot-pooled}) shows a mixed community structure consisting of relatively greater proportions of \textit{Synechococcus} especially in the northern hemisphere  
($\sim 51\%$ in Gradients cruises, and $11.9\%$ in the P16S\&N cruises) and eukaryotic phytoplankton ($\sim$45 \% with a standard deviation of $\sim$16 \%) such as picoeukaryotes, nanophytoplankton, and diatoms (Figure~\ref{fig:barplot-pooled}). In the higher latitude waters poleward of these frontal regions, there is a pronounced increase in eukaryotic phytoplankton abundance for bioclusters 11 - 16 ranging from about $14\%$ to $100 \%$ (Figure~\ref{fig:barplot-pooled}).
\par 

\subsubsection{Analyzing 3-dimensional biological provinces}
\label{sec:main-application-two-at-a-time}

Here we detail 3D-provinces derived by applying \texttt{bioprovince} to pooled datasets of two cruises at a time: P16S with P16N, and Gradients 2 with Gradients 3. This pairing was specifically chosen because the data for each pair were collected during the same time of year. The pooled P16S and P16N dataset represents a sparsely sampled, but geographically extensive, north-south Pacific Ocean transect. In contrast, the pooled Gradients 2 and Gradients 3 dataset provides a dense sampling of a more geographically compact area. These two distinct datasets offer a valuable opportunity to evaluate the respective merits and drawbacks of wide-scale, sparse sampling versus localized, dense sampling for the delineation of biological provinces.
\par

We can see in Figure~\ref{fig:novel-provinces-from-pairs-of-cruises} (Panel A1, A2) that our 3D-provinces line up well with existing Longhurst boundaries, while additionally revealing much finer spatial partitions across latitude and depth. Figure~\ref{fig:heatmap-compare-to-longhurst} summarizes the results using 2-way cross-tabulated tables across the two categories: (1) 3D-provinces as columns, and (2) Longhurst provinces as rows. Each row's measurements show that some Longhurst province match exactly with our novel 3D-province (e.g., ANTA which matches 3D-province 9 100\% from the paired P16S/P16N cruise) (Figure~\ref{fig:heatmap-compare-to-longhurst}). 

In the oligotrophic gyres, which are classically divided into only two Longhurst provinces (SPSG and NPTG), our \texttt{bioprovince} algorithm finds much more biologically distinct 3D-provinces. 
For example, Longhurst province SPSG is subdivided into three 3D-provinces 3, 4, and 5 (23.1\%, 23.1\% and 34.6\% respectively) in the upper euphotic zone, and 2 and 8 (15.4\% and 3.8\%) in the deeper euphotic zone. Similarly, Longhurst province NPTG consists of 3D-provinces 3, 4, and 5 (11.8\%, 17.6\% and 29.4\%) in the upper euphotic zone, and 2 and 8 (23.5\% and 17.6\%) in the deeper euphotic zone.
Our results also identify the upper euphotic zone of the equatorial Pacific (3D-province 1) as an important ecological boundary separating the oligotrophic regions of the N and S Pacific (Figure~\ref{fig:novel-provinces-from-pairs-of-cruises} panels A1 and B1). Interestingly, the classical Longhurst province PEQD does not specify depth-wise separation in equatorial waters, while our 3D-provinces clearly split this region into two -- one upper layer (3D-province 1, 53.3\%) and one deeper layer (3D-province 2, 46.7\%).
Moving poleward from oligotrophic gyres, our algorithm identified transitional 3D-provinces (6 and 7) separating those gyres in the upper 100 meters from higher-latitude 3D-provinces (9–11). These higher-latitude provinces exhibited greater taxonomic variability and relative abundance of eukaryotic phytoplankton (Panel B of Figure~\ref{fig:novel-provinces-from-pairs-of-cruises}). 

\par

%Also, deeper province boundaries generally coincided with the base of the euphotic zone, except in ANTA (3D-province 11), where no distinct deeper boundary was observed.
%\par

Finally, in Figure~\ref{fig:environment-isnt-enough} we overlay the biological and Longhurst classifications
on the abiotic conditions at the samples from the two cruises P16S \& P16N. We first see from Panel A that the samples are impossible to properly cluster with temperature and salinity alone, as no natural groups emerge that are distinguishable from others. Longhurst provinces explain away some of the variation in temperature and salinity -- as shown in Panel B. However, the biological clusters (shown as numbers) partition the samples much further than the Longhurst provinces (shown as shapes). Furthermore, Figure~\ref{fig:environment-isnt-enough} Panel C show the 3D-provinces in temperature-salinity dimensions, and Panel D shows it on the principal components' space of seven abiotic factors. In both plots, it is clear that (1) the abiotic conditions of the ocean alone are not adequate for finding biological clusters, and (2) the novel genomic metabarcoding data we are using is highly useful for defining new biological clusters and provinces.

\section{Discussion}

\subsection{Insights from a single-cruise application}
Our biological clusters and 3D-province estimates from the P16N cruise data (Figure~\ref{fig:workflow}) are consistent with studies of the same region that leveraged biogeochemical measurements to identify major differences in primary productivity across several known Longhurst provinces boundaries \citep{Barber1996}. Specifically, the algorithm reproduces the separation between oligotrophic gyres—such as the South Pacific Subtropical Gyre (SPSG) and North Pacific Tropical Gyre (NPTG)—and regions of elevated productivity, including (i) equatorial waters in the North Pacific Equatorial Countercurrent (PNEC) and Pacific Equatorial Divergence (PEQD) provinces, (ii) frontal regions like the North Pacific Polar Front (NPPF), and (iii) subpolar regions such as the Pacific Subarctic Gyres Province (East) (PSAE) (Figure~\ref{fig:workflow}). Beyond confirming these known patterns, the \texttt{bioprovince} approach reveals previously uncharacterized microbial community structure at depth, highlighting coherent vertical organization within each Longhurst province that is largely invisible in surface-only analyses (Figure~\ref{fig:workflow}). This deeper structure underscores the value of integrating high-resolution ASV data, demonstrating that biologically meaningful provinces extend beyond latitudinal productivity gradients and into the three-dimensional structure of the water column.

\subsection{Oceanographic drivers of phytoplankton community structure}
We extend the application of the \texttt{bioprovince} framework by pooling together five latitudinal cruises in the Pacific Ocean from the GRUMP dataset. The analysis revealed 16 distinct biological clusters that capture both surface and depth-resolved variation in phytoplankton communities (Figure~\ref{fig:pooled}). These findings support the coupling between physical, chemical, and biological processes that organize marine phytoplankton communities. For instance, continuous wind-driven upwelling in the equatorial Pacific enhances nutrient supply, supporting elevated new production in these high-nutrient, low-chlorophyll waters \citep{Barber1991Regulation, Dugdale1998Silicate}. In this region, we observe the recurring presence of biocluster 1 (i.e., P15S-01, P16N-01) across multiple cruises (Figure~\ref{fig:pooled}). Beneath the upper euphotic zone, another distinct biocluster (e.g., P16N-02) emerges, characterized by higher relative abundances of eukaryotic phytoplankton such as diatoms, prymnesiophytes, and pelagophytes, consistent with enhanced nutrient availability in the equatorial Pacific waters that separate two major oligotrophic gyres (Figure~\ref{fig:barplot-pooled}).  
\par

Poleward from the oligotrophic gyres, the decline of \textit{Prochlorococcus} and the rise of larger phytoplankton taxa, such as diatoms and bolidophytes, also coincide with increasing nutrient availability, marking additional transitions across environmental gradients that similarly support higher primary productivity \citep{Dutkiewicz2024Multiple}. Beyond shifts in environmental gradients, this collapse with the \textit{Prochlorococcus} populations has also been shown to be impacted by ecological interactions (e.g., shared predation, impact of viruses) \citep{Carlson2022, Follett2022}. Remarkably, the patterns of these bioclusters appear symmetrical across hemispheres and also extend into deeper waters (Figure~\ref{fig:pooled}). This suggests that at equivalent latitudes relative to the equator, similar bioclusters—and by extension, comparable ecological niches—occur in both the northern and southern Pacific. These observations further prove that the GRUMP dataset is highly consistent and comparable across different data collection methods in the same spatial latitude-depth area.
\par

Seasonal variability further influences biocluster assignments. For example, samples collected across comparable latitudes in the P15S and P16S cruises exhibit different biocluster memberships in similar Longhurst regions. Similarly, in the N. Pacific (i.e., NPTG, NPPF, PSAE) samples collected in the same region across three separate cruises (P16N-12, G2-12, G3-12) all belong to biocluster 12, but with P16N-12 located significantly more poleward than G2-12 and G3-12 -- highlighting seasonally {\it shifting} ecological niches within the NPTZ. In sum, we see in our analysis that when data collected in very different seasons are pooled together, the samples' estimated clusterings are often due not to purely spatial difference, but due to the seasonality difference. The two are hard to disentangle and without significant postprocessing to post-hoc combine certain biocluster memberships, it is challenging to use the bioprovince algorithm for geographically coherent and interpretable 3D-provinces. Thus, we recommend defining 3D-provinces only from data pooled from cruises taken at a similar time of year. In Section~\ref{sec:main-application-two-at-a-time}, we do precisely this and discuss the results in the following section.
\par 

\subsection{Ecological insights from 3-dimensional provinces}
By combining data across paired cruises sampled during similar times of the year, the \texttt{bioprovince} algorithm resolves biologically coherent 3D-provinces that extend the classic two-dimensional view of ocean biogeography. For example, within the oligotrophic gyres, since both the oligotrophic NPTG and ultra-oligotrophic SPSG exhibit very low chlorophyll-\textit{a} concentrations with very high abundances of \textit{Prochlorococcus} \citep{longhurst-book, Morel2010, Flombaum2013Present}, the Longhurst framework delineates only two broad provinces. However, our method identifies multiple, vertically distinct 3D-provinces corresponding to unique ecological niches  (Figure~\ref{fig:novel-provinces-from-pairs-of-cruises}). This finer-scale partitioning is largely due to the granular, ASV-resolution survey of plankton in the GRUMP dataset which captures well-known niche differentiation among \textit{Prochlorococcus} ecotypes. This includes: (i) high-light–adapted clades dominating the upper euphotic zone, while low-light–adapted ecotypes prevail at depth and (ii) a latitudinal transition from HLII to HLI ecotypes moving poleward towards cooler waters from the warmer, central oligotrophic gyres (Supp. Figure~\ref{fig:pro-ecotype}). This separation of \textit{Prochlorococcus} ecotypes by light and temperature regimes has been well established using empirical observations in other ocean basins and laboratory-based culture studies \citep{johnson2006niche, biller2015prochlorococcus}.
\par

The deeper 3D-province boundaries (in Figure~\ref{fig:novel-provinces-from-pairs-of-cruises}) are largely congruent with the base of the euphotic zone, which further demonstrates the controlling role of abiotic factors such as light availability in structuring phytoplankton communities (Figure~\ref{fig:novel-provinces-from-pairs-of-cruises} A1). One notable exception is the absence of a deep provincial boundary in the Antarctic region, suggesting greater homogeneity in phytoplankton distribution throughout the water column that is likely driven by rapid particle sinking and strong vertical mixing \citep{smith2015vertical, zuniga2021sinking}. This vertical homogeneity in phytoplankton community structure, dominated by diatoms and prymensiophytes (Figure \ref{fig:novel-provinces-from-pairs-of-cruises} B1), also illuminates the potential sensitivity of 3D-province boundaries to taxonomic scope. For instance, the inclusion of mixotrophic taxa such as dinoflagellates could reveal additional vertical transitions, as these organisms can persist across light and nutrient gradients by switching trophic strategies \citep{Cohen2021Dinoflagellates}. Overall, the 3D-provinces derived here highlight how integrating biological and environmental information can reveal complex ecological zonation in the ocean interior.
\par

\subsection{Comparison with conventional biogeographical frameworks}
Our 3D-provinces exhibit strong concordance with established frameworks such as the Longhurst provinces, while simultaneously capturing finer vertical and latitudinal variation. Cross-tabulation analyses confirm that conventional Longhurst provinces are largely embedded within the new \texttt{bioprovinces} (Figure~\ref{fig:heatmap-compare-to-longhurst}), underscoring that physical and chemical properties continue to structure large-scale marine biogeography. Notably, the near one-to-one correspondence between the Antarctic Longhurst province (ANTA) and 3D-province 11, as well as between two polar Longhurst provinces (PSAE and SANT) with 3D-bioprovince 9 and 10 in shallow waters, highlights that fundamental ecological boundaries are preserved even when defined via genomic data.
\par

Going beyond simply validating the classical province framework, the \texttt{bioprovince} approach further reveals biologically meaningful refinements. In the North Pacific Transition Zone, the boundaries inferred from Gradients 2 and 3 data are displaced southward relative to the canonical NPTG/NPPF/PSAE divisions, likely reflecting seasonal shifts in frontal positions. We can further examine similar oceanographic features in the S. Pacific -- specifically the Polar Front (PF), subtropical front (STF), and equatorial upwelling (EQ). These dynamic regions are well-studied by \cite{Raes2018} (based on analyzing the densely sampled P15S cruise) as persistent {\it ecological} boundaries. Indeed, in our cruise-by-cruise data analysis (in Appendix Figure~\ref{fig:all-cruises}), novel 3D-province boundaries successfully separate distinct phytoplankton communities across all three regions (PF, STF, and EQ). This successful delineation is evident not only in the P15S cruise but also in the P16S and P16N cruises, which sampled the region approximately a decade earlier. By revealing 3D-province boundaries that persist across datasets collected across a long time period, our method corroborates findings in \cite{Raes2018} regarding the importance of persistent oceanographic features as ecological boundaries.  
\par

Our methodology successfully utilizes deeper water abiotic conditions to extend the conventionally known province definition beyond surface-level delineation by revealing new coherent biological niches of deeper-water marine microbes -- especially in the oligotrophic and equatorial waters. Additionally, it is clear the abiotic conditions alone would not allow for groupings made using biological conditions, as depicted in Figure~\ref{fig:environment-isnt-enough}. Collectively, these observations demonstrate that the \texttt{bioprovince} framework reconciles biological variability with physical and chemical oceanographic structure. By providing a data-driven, vertically resolved, and temporally robust view of marine ecological patterns, this approach complements and extends conventional frameworks, enabling a more comprehensive understanding of the factors shaping marine biodiversity across space and depth.
\par

\section{Conclusion}
The ability to define biologically coherent 3D-provinces has important implications for ocean ecology and climate science. Our \texttt{bioprovince} algorithm provides a reproducible and objective means to partition the ocean into ecologically meaningful regions, which can improve the resolution and accuracy of biogeochemical and ecosystem models. By incorporating biological information directly, \texttt{bioprovince} enhances our understanding of how productivity, nutrient cycling, and microbial community structure vary through space and depth.
\par

To further enhance the utility of this framework, a crucial next step is to develop a statistical model for microbial ASV compositions observed across both time and space that enables principled statistical inference of estimated cluster memberships. This model should explicitly account for the annual seasonal cycle of the ocean, which our \texttt{biocluster} analysis revealed to be one of the main drivers of community composition.  Properly handling this temporal dependence in a statistical model will also allow for more robust out-of-sample province predictions in spatial locations farther away from existing samples, and in different times of the year. In addition, such a statistical model may also include a regression component to explicitly use environmental factors (e.g., temperature, salinity, nutrients), which could further enhance its prediction ability at new locations.
\par

Beyond phytoplankton, this framework can be readily extended to other microbial groups within the GRUMP dataset or applied to independent datasets such as the Atlantic Meridional Transect (AMT), bioGEOTRACES, or TARA Oceans. Temporal comparisons across historical and modern metabarcoding datasets could further illuminate how marine provinces shift seasonally or in response to climate variability. The \texttt{bioprovince} algorithm thus offers a generalizable approach for investigating spatial and temporal evolution in marine microbial biogeography, providing a foundation for integrating molecular observations into global oceanographic research.
\par

\section*{Acknowledgments}
This work was supported through grants by the Simons Collaboration on Computational Biogeochemical Modeling of Marine Ecosystems/CBIOMES (Grant ID: 1195553 to S.H.). This work was also supported by the Simons Foundation (CBIOMES-549943 to J.A.F.) and the National Science Foundation (NSF-OCE-1737409 to J.A.F.).

% Done using \usepackage[figuresonly]{endfloat}

%% file: data-statement.tex
\section*{Data Accessibility }

The GRUMP data is available at Simons CMAP (\url{https://simonscmap.com/catalog/datasets/GRUMP}).

%Raw sequences The forward and reverse sequences were submitted to the NCBI database under accession numbers GA02/ GA10: PRJNA1194189, GA03: PRJNA1194192, GP13: PRJNA1195113, SCOPE-Gradients 2: PRJNA1195115, SCOPE -Gradients 3: PRJNA1196422, P16N: PRJNA1196483, P16S: PRJNA1196490, P15S: PRJNA1196498, HEOBI: PRJNA1196504, IND-2017: PRJNA1196513, K-AXIS: PRJNA119651, POTATOE: PRJNA1198072, I08S: PRJNA1198088, I09N: PRJNA1198607, MOSAiC: PRJNA1198992, and FRAM: PRJNA1199044. 

%\paragraph{\bf Benefits Generated:}
%(this is a borrowed example from their page) A research collaboration was developed with scientists from the countries providing genetic samples, all collaborators are included as co-authors, the results of research have been shared with the provider communities and the broader scientific community (see above), and the research addresses a priority concern, in this case the conservation of organisms being studied. More broadly, our group is committed to international scientific partnerships, as well as institutional capacity building.

%% file: author-contribution.tex
\section*{Author Contributions}

Rafael Catoia and Sangwon Hyun designed the analytical pipeline, performed numerical experiments, and wrote the statistical software accompanying this manuscript. Nathan Williams bioinformatically prepared ASV data and also provided contextual information assisting with the use of this data. Yubin Raut assisted with generating the ASV data, provided feedback on the statistical framework and its applications to the ASV data, and contributed to data interpreation. Jed Fuhrman supervised the data curation and advised the ecological interpretation. All authors contributed substantially to writing the paper.

%% file: supp-content.tex
\setcounter{section}{0}

\section{Supplement: Additional figures}

\begin{figure}[ht!]
\centering
\includegraphics[width=.4\linewidth]{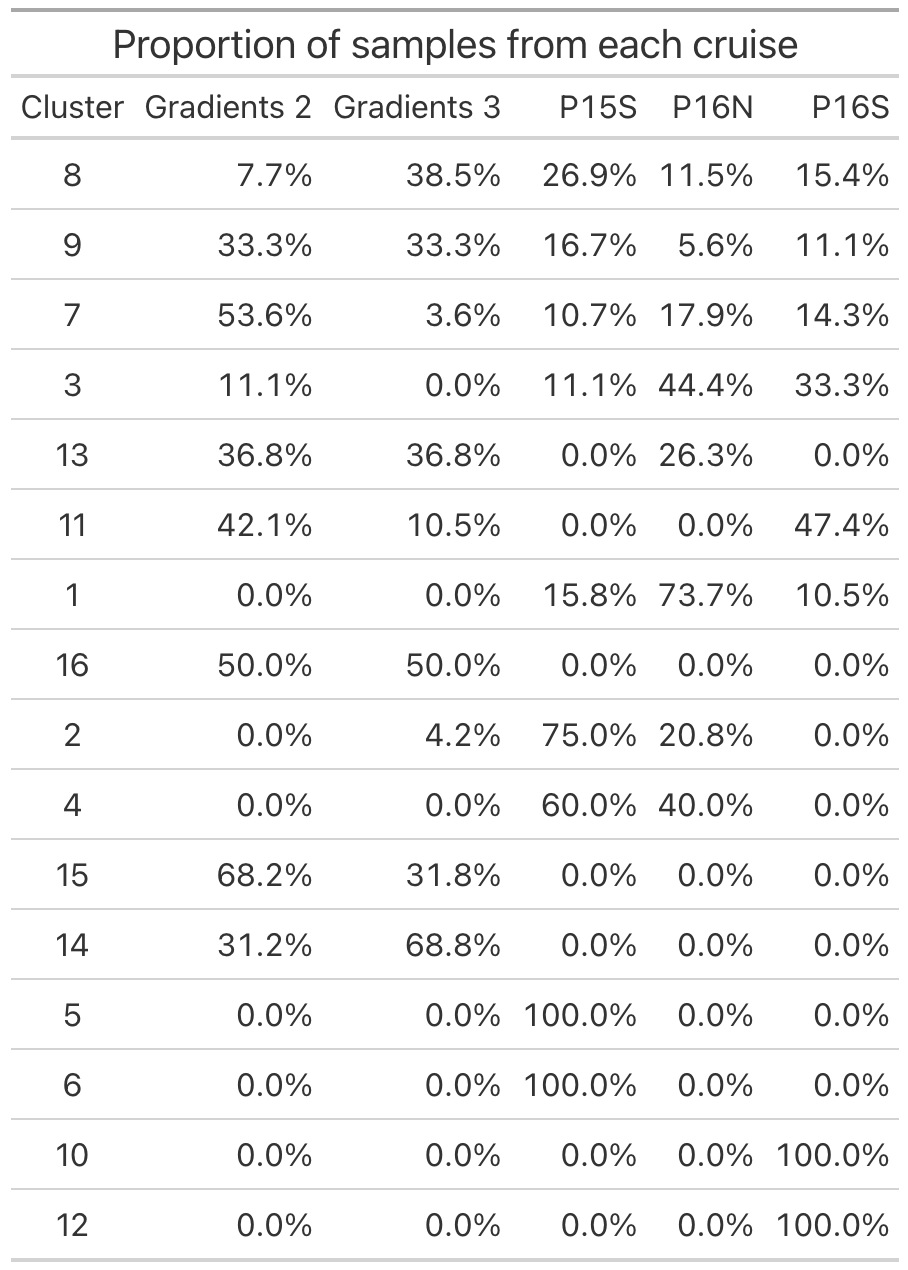}
\caption{\it The mix of clusters from each cruise in the 16-cluster analysis of Figure~\ref{fig:pooled} in Section~\ref{sec:main-application-pooled}. The table is sorted from top to bottom in order of how highly homogenous (mixed) those clusters' samples are across the five cruises. The homogeneity is measured using a KL divergence against an equal mix of five cruises. Here, 12 out of the 16 clusters are found are from {\it multiple} cruises -- that is, those clusters consist of at least a 70\%-30\% mix (of samples) from two cruises. Only clusters 13, 14, 15, and 16 (all in Longhurst SANT and ANTA)  originate entirely from a single cruise.}
\label{tab:pooled-mix}
\end{figure}

\begin{figure}[ht!]
\centering
\includegraphics[width = .95\linewidth]{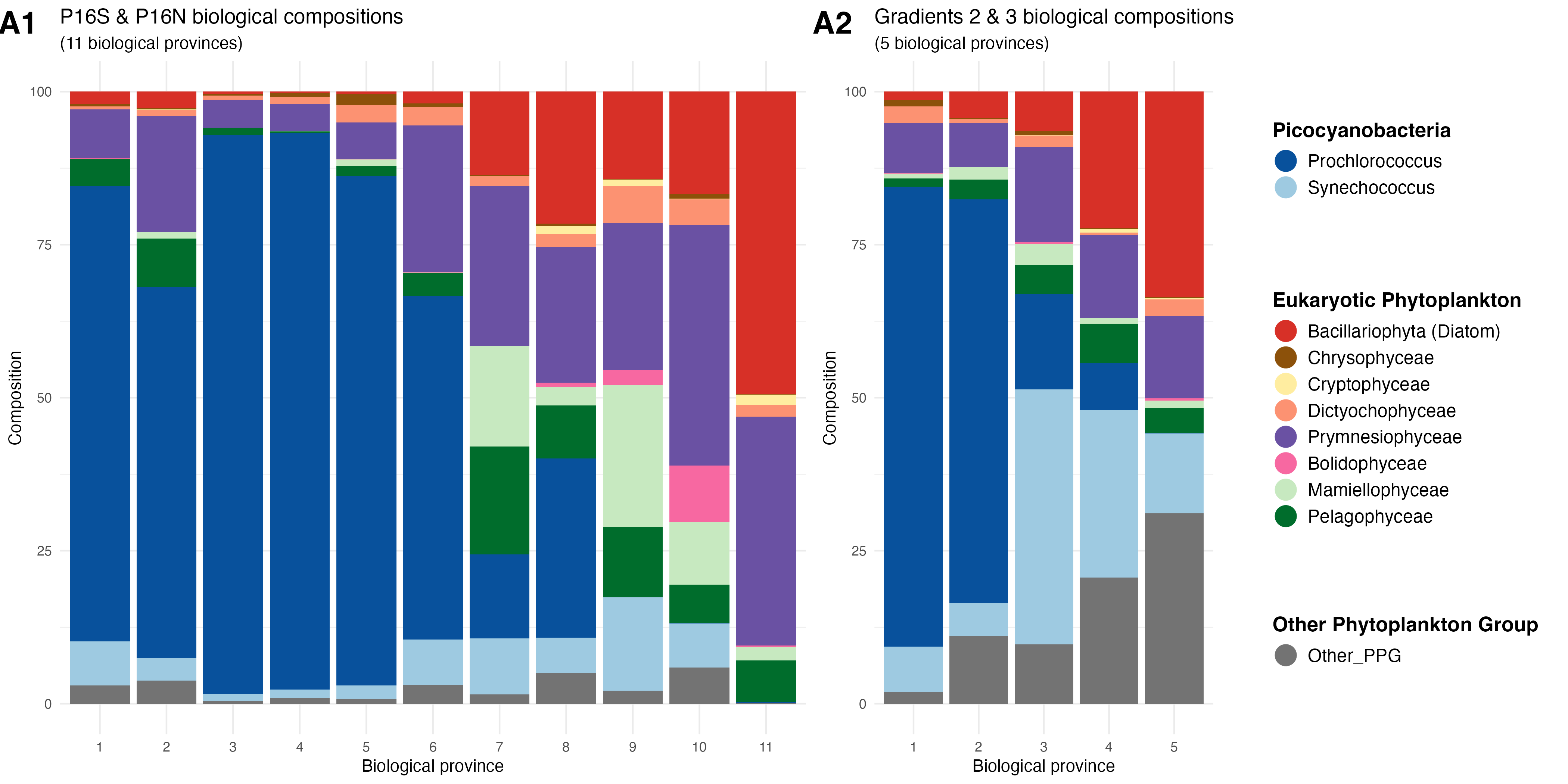}
\includegraphics[width = .62\linewidth]{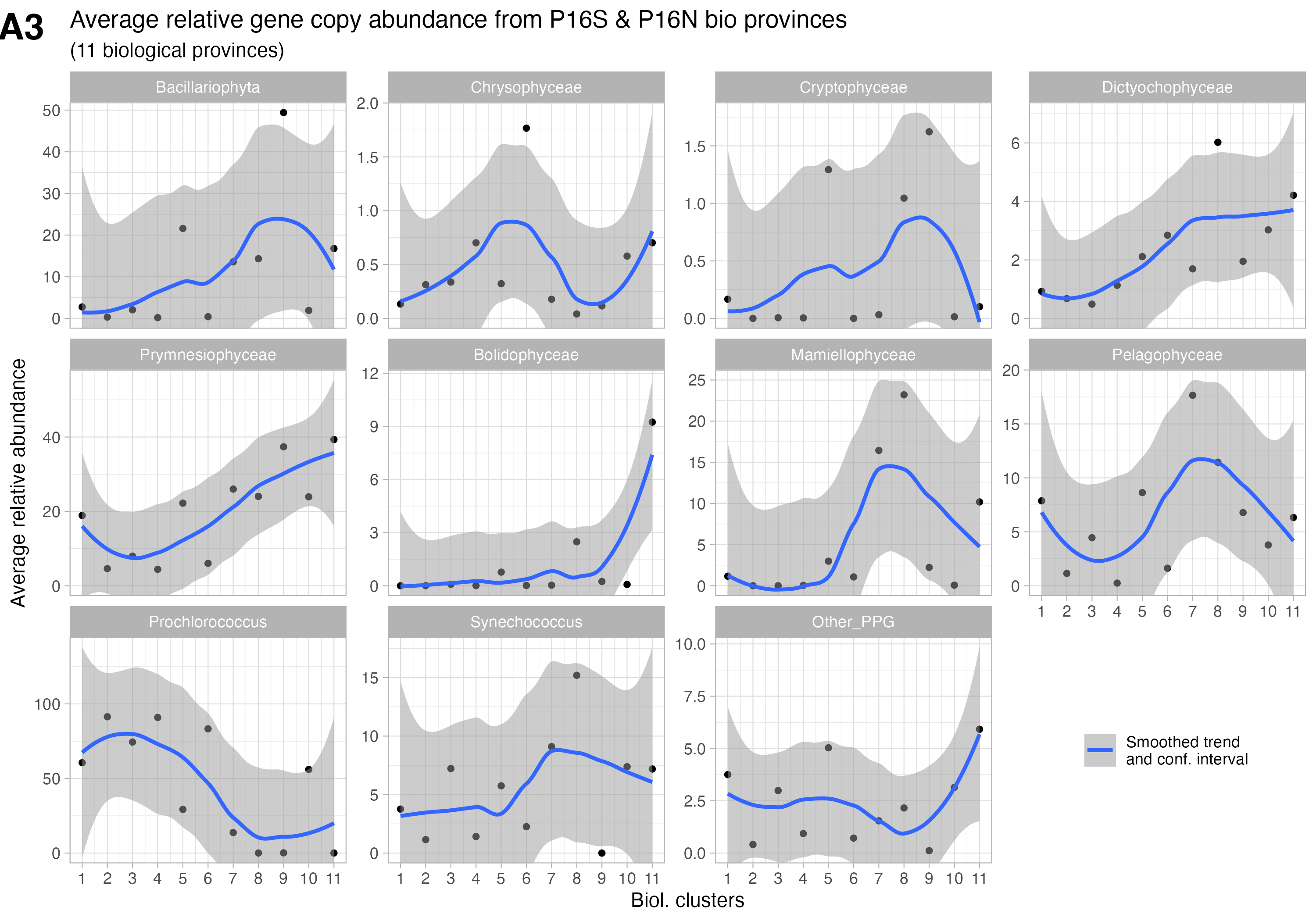}
\includegraphics[width = .37\linewidth]{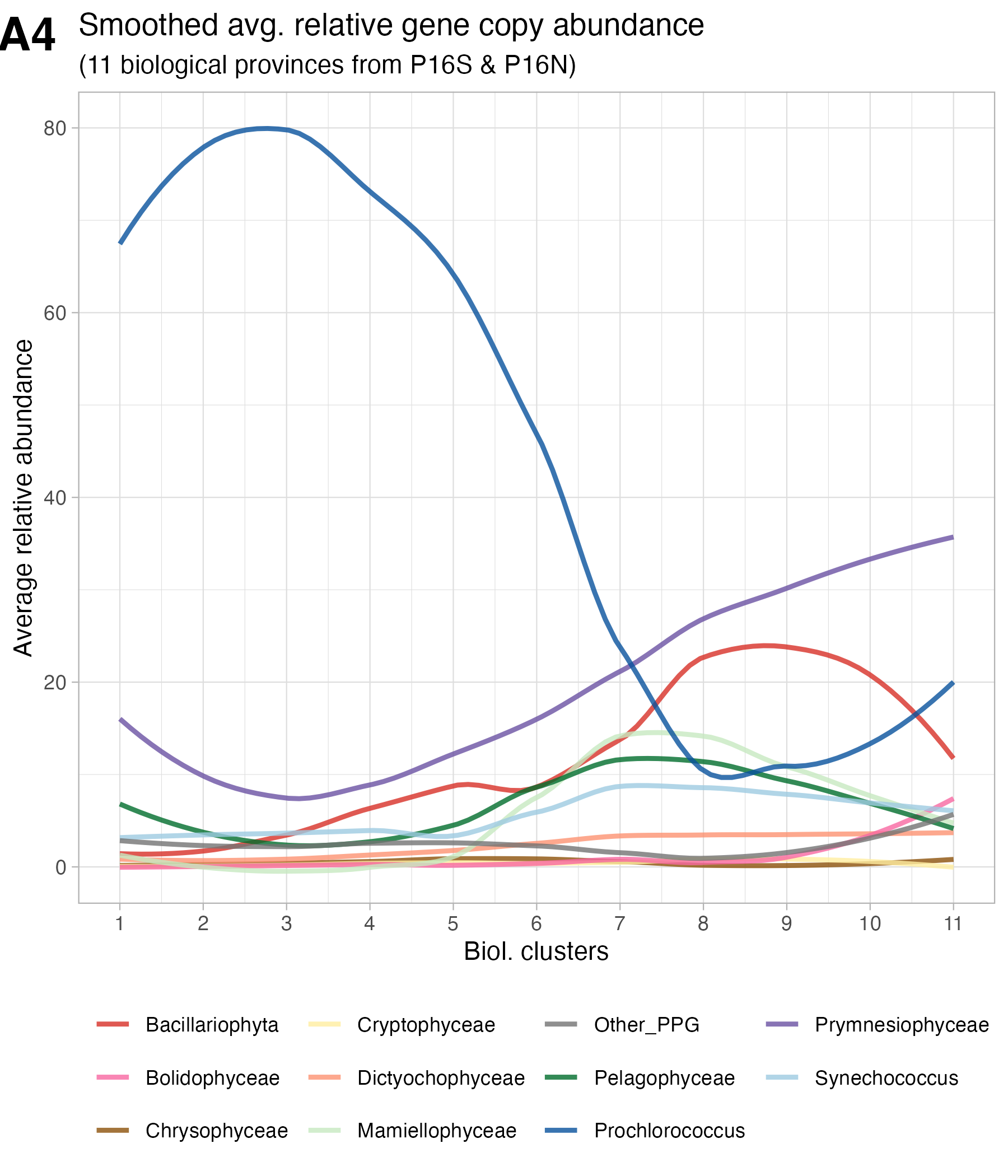}
    \caption{\it  The average relative composition for each bioprovince, at the ecologically relevant plankton group level, is shown as barplots (top row) and line plots (bottom). This figure is supplementary to Figure~\ref{fig:novel-provinces-from-pairs-of-cruises}.}
\label{fig:novel-provinces-from-pairs-of-cruises-barplots}
\end{figure}

Additional figures (Figure~\ref{tab:pooled-mix} and \ref{fig:novel-provinces-from-pairs-of-cruises-barplots}) are shown here.

%\subsection{Appendix: Separate bioprovince application to five transect cruises}

% Applying our bioprovince algorithm in several different ways, we highlight important newly discovered biological provinces along the Pacific ocean transect, comparisons to conventional province definitions, and ecological interpretations that result from delineating the new provinces.

%\paragraph{Analyzing new biological provinces}
%Next, we compare our estimated biological provinces from five cruises that traveled the northern part of the transect (Gradients 2 and 3 and P16N), and the southern part of transect (P15S vs P16N). Figure~\ref{fig:all-cruises} show the results of applying {\tt bioprovince} to all cruises, lined up with one another by latitude.

\begin{figure}
\centering
\includegraphics[width = \linewidth]{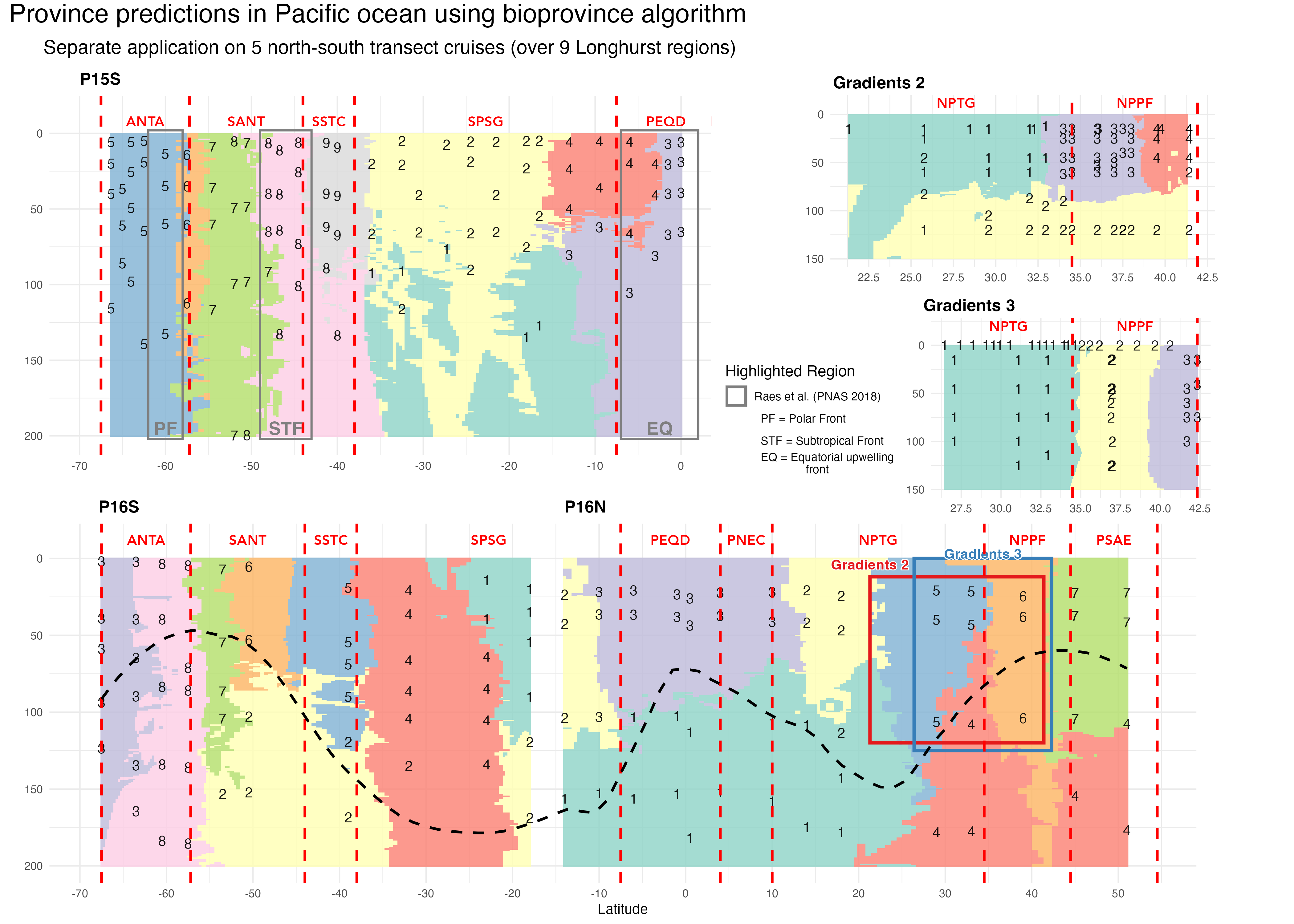}
    \caption{\it Comparing the estimated provinces computed separately from the five transect cruises. Colors show the estimated bio-province, and the numbers show the cluster memberships estimated at the sample locations. Longhurst boundaries are shown in vertical red lines.  The top-left figure (P15S) also shows bio-diversity fronts from \cite{Raes2018}. The bioprovince estimates in Figure~\ref{fig:workflow} and the current Figure~\ref{fig:all-cruises} are consistent with several well-established boundaries between Longhurst provinces that separate oligotrophic gyres (e.g., SPSG, NPTG) from regions of elevated primary productivity, including equatorial waters (e.g., PEQD, PNEC), frontal zones (e.g., NPPF, SSTC), and subpolar and polar regions (e.g., PSAE, SANT, ANTA).
    }
\label{fig:all-cruises}
\end{figure}

%\begin{figure}
%[width = \linewidth]{appendix-figures/environmental-conditions-draft.png}
%\includegraphics[width = \linewidth]{appendix-figures/barplots.png}
%\caption{\it The average
%\notate{SH, YR: different samples have different sampling depths, which might confound? YR with check with Nathan \& Jed.} \notate{SH: need to update panel C with new clusters/provinces from the latest (final) round of analysis.}}
%\label{fig:all-cruises-ecological-interpretation}
%\end{figure}

\begin{figure}
\centering
\includegraphics[width = .8\linewidth]{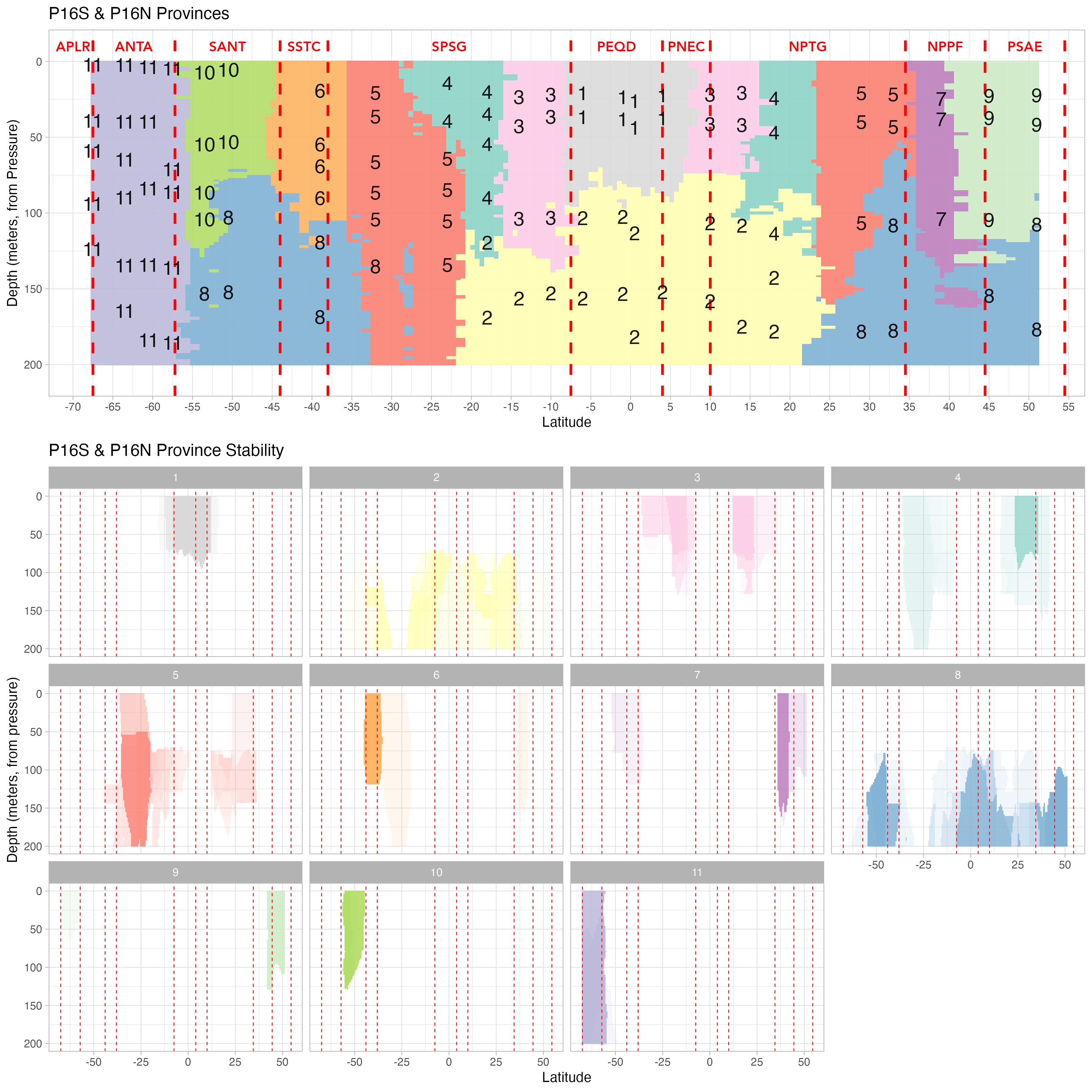}
\caption{\it Plotting the stability \ref{subsec:stability} of province estimates for P16 cruise data. The top panel reproduces the provinces estimated from data from the pair for cruises (P16S and P16N), and the bottom plot shows the stability estimates of each 3d-bioprovince in separate panels, with opacity of the color proportional to the stability. Notably, the surface provinces appear more stable, while the two deeper provinces are less stable, especially with respect to the province estimates in the top panel.}
%. In shallower depths along latitude, the North Pacific Gyre province is much stabler than in other latitudes, which points to a more rapid change in the environmental gradient.}
\label{fig:stability-p16}
\end{figure}

\begin{figure}
\centering
\includegraphics[width = .7\linewidth]{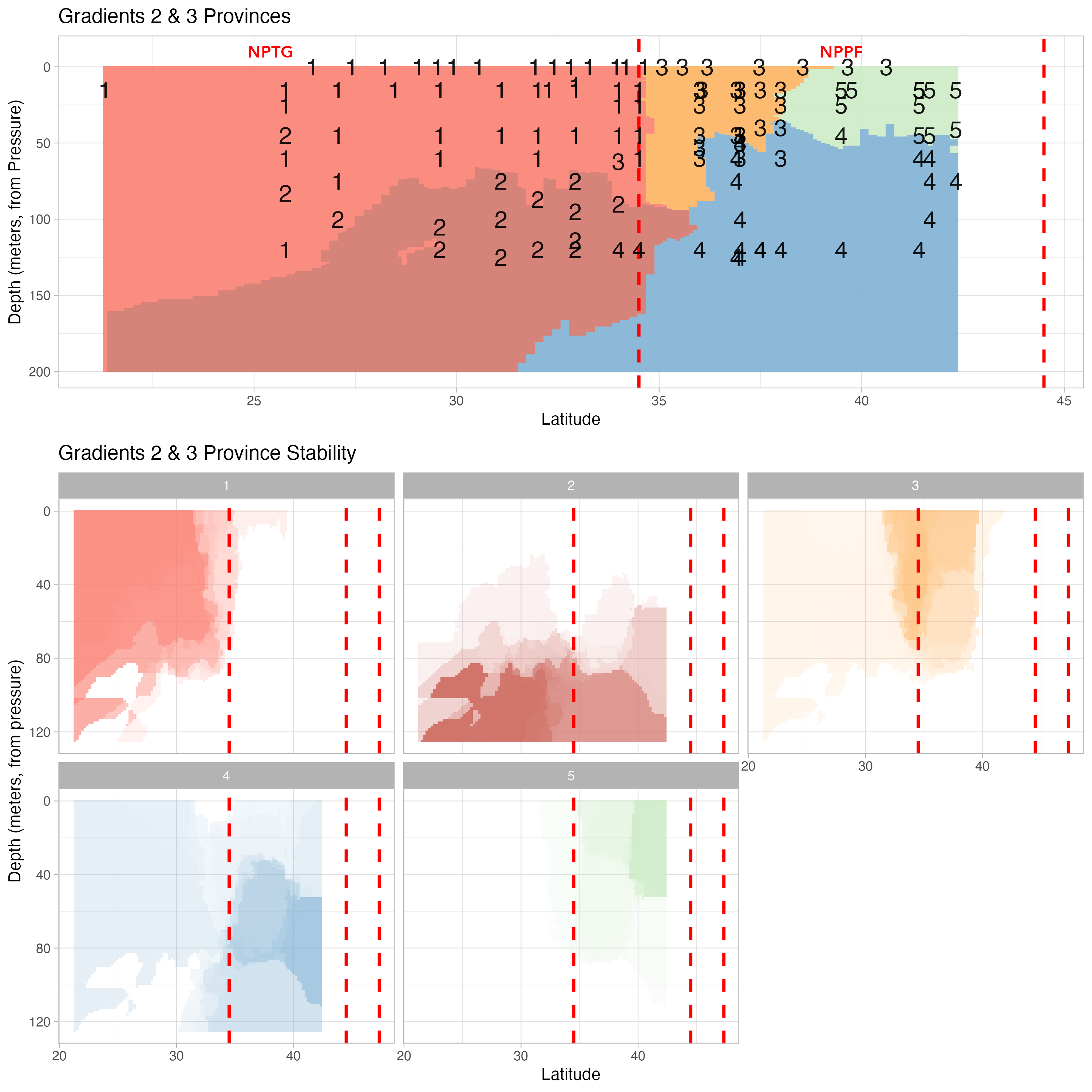}
\caption{\it 
Plotting the stability of the province predictions \ref{subsec:stability} for Gradients cruise data. The top panel reproduces the provinces estimated from data from the pair of cruises Gradients 2 \& 3, and the bottom plot shows the stability estimates of the province predictions in separate panels, with opacity of the color proportional to the stability.}
%. In shallower depths along latitude, the North Pacific Gyre province is much stabler than in other latitudes, which points to a more rapid change in the environmental gradient.}
\label{fig:stability-gradients}
\end{figure}

%\paragraph{New provinces' ecological interpretation}
%
%Figure~\ref{fig:all-cruises-ecological-interpretation} shows the average
%compositions and the environmental conditions in each estimated province.
%Clearly, our estimated biological provinces are heavily overlapping in terms of
%environmental factors, and clustering or province estimates using these
%physical/chemical factors would have produced very different, probably more
%simplistic province boundaries.

\section{Supplement: Full procedure for choosing $r$}
\label{sec:supp-choose-r}

We took all pairs of distances between the $n^{(l)}$ samples in the data for a single cruise ($l$) that are close enough to be relevant. (In our analysis, we isolate our attention to latitudes that are within 3 degrees, and depths that are within 10 meters.) Then we conducted a linear regression of (1) the difference in Aitchison distance onto (2) the latitude (degrees). Call this regression model: $d_{\text{bio}} = b_{01}^{(l)} + b_1^{(l)} \cdot d_{\text{lat}}$. The slope $b_1$ quantifies the degree of \textit{decay} in biological similarity (as measured by Aitchison distance) as one moves away in latitude. Next, we conduct another linear regression of (1) the difference in Aitchison distance onto (2) depth (in meters). Call this regression: $d_{\text{bio}} = b_{02}^{(l)} + b_2^{(l)} \cdot d_{\text{depth}}$.
Call $b_1 = \frac{\sum_{l=1}^5 b_1^{(l)} n^{(l)}}{\sum_{l=1}^5 n^{(l)}}$ the weighted average of the regression slopes of latitude; likewise, call $b_2 = \frac{\sum_{l=1}^5 b_2^{(l)} n^{(l)}}{\sum_{l=1}^5 n^{(l)}}$ the weighted average of regression slopes of depth.

In order to create a new rescaled version of latitude (call this scaled\_lat) which
has the same footing as $d_{\text{depth}}$, we multiplied latitude by a factor of $r =b_1/b_2$ to 
 obtain the scaled latitude:
\begin{equation}
\label{eq:scaled-lat}
\text{scaled\_lat} = \text{lat} \cdot \frac{b_1}{b_2},
\end{equation}
or equivalently, $\text{lat} = \text{scaled\_lat} \cdot \frac{b_2}{b_1}$. Lastly, we see how $r = b_1/b_2$ should come into play in the spatial distance between, say, two samples $i_1$ and $i_2$:
$$\sqrt{ (r \cdot \text{lat}(i_1) - r \cdot \text{lat}(i_2))^2 + (\text{depth}(i_1) - \text{depth}(i_2))^2},$$ 
which \textit{upweights} latitude by a factor of $r$, plays the role of rescaling latitude to be equivalent
to depth. 

\begin{figure}
\centering
\includegraphics[width=\linewidth]{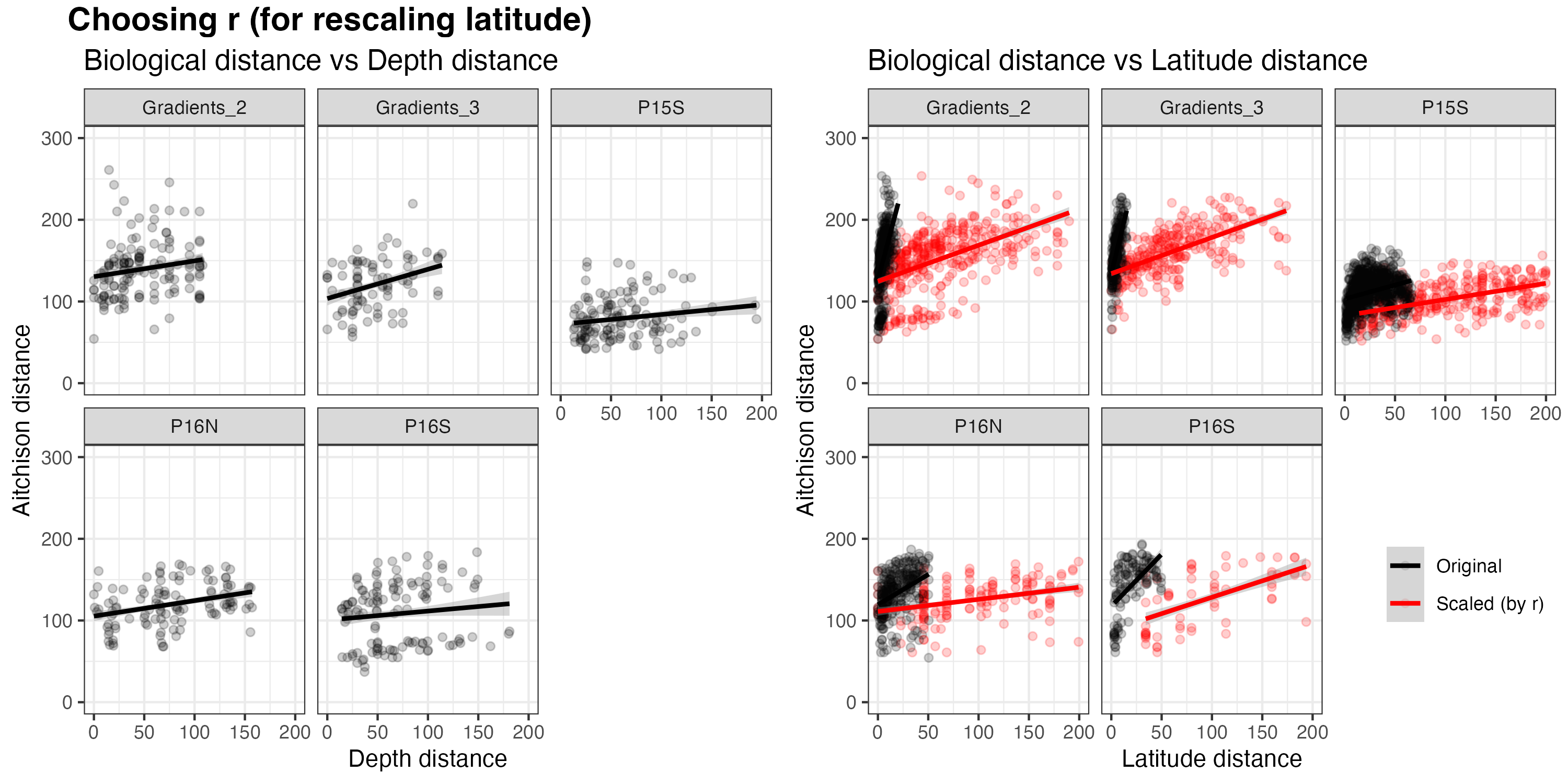}
\caption{\it The left panel shows a scatterplot of the pairwise biological distances (measured by Aitchison distance of the compositions at each pair of samples) between samples (y axis) over the depth distance between samples (x axis) using black points, as well as a linear regression line. Each subpanel shows this analysis from one cruise. The slope of the regression approximately measures the speed of \textit{decay} in the biological similarity as one moves farther away in depth. Next, the right-hand-side shows a scatterplot of the biological distances and the latitude distance in black points, with a linear regression line. The red points, show the scaled distances $\text{scaled\_lat}$ from \eqref{eq:scaled-lat} as x values instead, and the corresponding linear regression of biological distances on \text{scaled\_lat}. The red slopes are comparable in size to the left panel's slopes, which verifies visually that the rescaling of latitude is sensible.}
\label{fig:choose-r}
\end{figure}

Figure~\ref{fig:choose-r} shows an example of the two types of regressions on each cruise. Here, the red slopes of the regression of $d_{\text{bio}}$ onto $\text{scaled\_lat}$ in the right panel are similar in scale to the slopes of the regression lines of $d_{\text{bio}}$ onto $\text{depth}$ in the left panel. This visually verifies that the rescaling of latitude results in a similar rate of decay in biological similarity over both rescaled latitude and depth.

\section{Supplement: Full procedure for choosing $\alpha$}
\label{sec:supp-choose-alpha}

First, augment the compositional dataset $P$ with four rows that are randomly selected four compositions from the existing dataset $P$. Then, form a scaled $(n+4) \times (n+4)$ pairwise Aitchison distance matrix $D_{\text{bio}}^{\text{aug}}$ of the augmented (n+4) compositions as in \eqref{eq:aitchison-mat}. Also augment the latitude and longitude by four points that are placed at the four corners in latitude and depth. Then, calculate the scaled $(n+4) \times (n+4)$ distance matrix as in \eqref{eq:spatial-mat}, $D_{\text{spatial}}^{\text{aug}}$.
After scaling both $D_{\text{bio}}^{\text{aug}}$ and $D_{\text{spatial}}^{\text{aug}}$, take an entry-wise convex combination as in \eqref{eq:mix-matrix} to form $D^{\text{aug}}$. Next, using this matrix, form a 2-dimensional multidimensional scaling (MDS) plot, and count the relative number of the points in the MDS coordinate space that are in the {\it convex hull} connecting those four points. 

Repeat this $L=100$ times to calculate this score $n_l(\alpha)$ for the $l$'th replicate. Also call the {\it null} scores as $n^{\text{null}}_l(\alpha)$, in which the four corners are also randomly selected spatial points in the sample. Null scores are understood to be the score values when the four additional data points have no spatial meaning. Finally, pick $\alpha$ to be a value that does not exceed the first value where the two $\{n_l(\alpha), l=1,\cdots, 100\}$ and $\{n_l^{\text{null}}(\alpha), l=1,\cdots, 100\}$ are visibly separated. The first panel in Figure~\ref{fig:workflow} shows an example of the ``non-null" scores $n_l(\alpha)$ and null scores $n^{\text{null}}(\alpha)$ from 
the example in Section~\ref{sec:one-cruise-application} -- with a 95\% balanced quantile bands shown as solid lines. This plot suggests that $\alpha$ values that are smaller or equal to $0.4$ are values for which the spatial information ($D_\text{geo}$) has not fully {\it saturated} the clustering, since there was no clear separation between the scores and the null scores for $\alpha \le 0.4$.

%\section{Appendix: Additional figures}
%
%\begin{figure}[ht!]
%\centering
%\includegraphics[width = .7\linewidth]{appendix-figures/choose-k.png}
%\caption{\it Placeholder.}
%\label{fig:choose-K}
%\end{figure}
%
%
%\begin{figure}[ht!]
%\centering
%\includegraphics[width = \linewidth]{appendix-figures/mds-for-tuning-alpha.png}
%\caption{\it Placeholder.}
%\label{fig:mds-for-tuning-alpha}
%\end{figure}
%
%\begin{figure}[ht!]
%\centering
%\includegraphics[width = .5\linewidth]{appendix-figures/score-for-tuning-alpha.png}
%\caption{\it (Draft figure) Example of null and non-null scores for choosing $\alpha$, as described in Appendix Figure~\ref{sec:supp-choose-alpha}.}
%\label{fig:score-for-tuning-alpha}
%\end{figure}
%

%\section{Prochlorococcus ecotype distribution along P16S \& P16N cruises}
\begin{figure}[ht!]
\centering
\includegraphics[width = \linewidth]{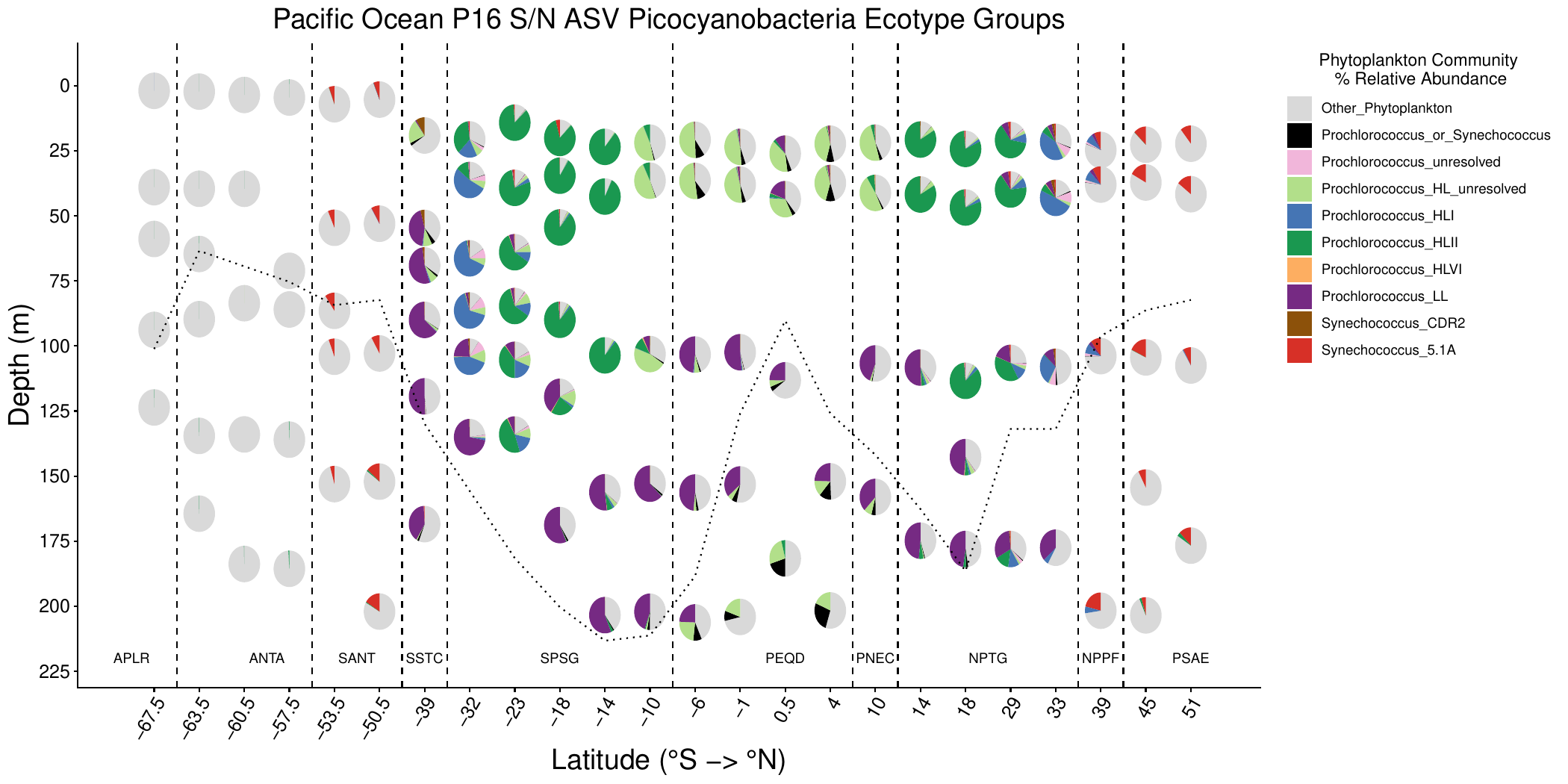}
\caption{\it Vertical and spatial distribution of phytoplankton community composition across the P16 S/N transect. Each pie chart represents the aggregated relative abundance of ASVs belonging to major picocyanobacteria ecotypes in context to the residual phytoplankton ASVs including diazotrophic cyanobacteria and eukaryotic phytoplankton (derived from chloroplast 16S) colored in gray at a given sampling depth and latitude in the upper 200m along the transect. The dotted horizontal line within each station denotes the approximate lower boundary of the euphotic zone. Vertical dashed lines separate distinct biogeographical regions along the transect, annotated by Longhurst provinces.}
\label{fig:pro-ecotype}
\end{figure}

%\section{Temporary text about Figure 10}
%
%The bioprovince estimates in Figure~\ref{fig:workflow}  and Figure~\ref{fig:all-cruises} are consistent with several well-established boundaries between Longhurst provinces that separate oligotrophic gyres (e.g., SPSG, NPTG) from regions of elevated primary productivity, including equatorial waters (e.g., PEQD, PNEC), frontal zones (e.g., NPPF, SSTC), and subpolar and polar regions (e.g., PSAE, SANT, ANTA) 

%% file: main.bib
@article{RN35,

author = {Chavent, Marie and Kuentz-Simonet, Vanessa and Labenne, Amaury and Saracco, Jérôme and Chavent, Marie and Kuentz-Simonet, Vanessa and Labenne, Amaury and Saracco, Jérôme},

title = {ClustGeo: an R package for hierarchical clustering with spatial constraints},

journal = {Computational Statistics 2018 33:4},

volume = {33},

number = {4},

ISSN = {1613-9658},

DOI = {10.1007/s00180-018-0791-1},

url = {https://link.springer.com/article/10.1007/s00180-018-0791-1},

year = {2018-01-20},

type = {Journal Article}

}

@article{palomares2020fishery,
  title={Fishery biomass trends of exploited fish populations in marine ecoregions, climatic zones and ocean basins},
  author={Palomares, MLD and Froese, Rainer and Derrick, B and Meeuwig, JJ and N{\"o}el, S-L and Tsui, G and Woroniak, J and Zeller, D and Pauly, D},
  journal={Estuarine, Coastal and Shelf Science},
  volume={243},
  pages={106896},
  year={2020},
  publisher={Elsevier}
}

@article{parada2016every,
  title={Every base matters: assessing small subunit rRNA primers for marine microbiomes with mock communities, time series and global field samples},
  author={Parada, Alma E and Needham, David M and Fuhrman, Jed A},
  journal={Environmental microbiology},
  volume={18},
  number={5},
  pages={1403--1414},
  year={2016},
  publisher={Wiley Online Library}
}

@article{Plankton-Distribution,
    author = {Winter, Jordan AND Hynes, Annette AND Berthiaume, Chris AND Cain, Kelsy AND Armbrust, E. Virginia AND Ribalet, François},
 title = {Shifts in phytoplankton community structure across oceanic boundaries},   
journal = {PLOS ONE},
    publisher = {Public Library of Science},
    
    year = {2025},
    month = {06},
    volume = {20},
    url = {https://doi.org/10.1371/journal.pone.0324466},
    pages = {1-16},
     doi = {10.1371/journal.pone.0324466},
    number = {6},

}

@book{longhurst-book,
author = {Longhurst, A.R.},
year = {2007},
month = {01},
pages = {},
title = {Ecological Geography of the Sea},
doi = {10.1016/B978-0-12-455521-1.X5000-1},
publisher = {Academic Press}
}

@article{aitchison1982,
    author = {Aitchison, J.},
    title = {The Statistical Analysis of Compositional Data},
    journal = {Journal of the Royal Statistical Society: Series B (Methodological)},
    volume = {44},
    number = {2},
    pages = {139-160},
    year = {1982},
    month = {01},
    issn = {0035-9246},
    doi = {10.1111/j.2517-6161.1982.tb01195.x},
    url = {https://doi.org/10.1111/j.2517-6161.1982.tb01195.x},
    eprint = {https://academic.oup.com/jrsssb/article-pdf/44/2/139/49097906/jrsssb\_44\_2\_139.pdf},
}

@article{Ward1963,
  title = {Hierarchical Grouping to Optimize an Objective Function},
  volume = {58},
  ISSN = {1537-274X},
  url = {http://dx.doi.org/10.1080/01621459.1963.10500845},
  DOI = {10.1080/01621459.1963.10500845},
  number = {301},
  journal = {Journal of the American Statistical Association},
  publisher = {Informa UK Limited},
  author = {Ward,  Joe H.},
  year = {1963},
  month = mar,
  pages = {236–244}
}

@article{Nekola1999,
  title = {The distance decay of similarity in biogeography and ecology},
  volume = {26},
  ISSN = {1365-2699},
  url = {http://dx.doi.org/10.1046/j.1365-2699.1999.00305.x},
  DOI = {10.1046/j.1365-2699.1999.00305.x},
  number = {4},
  journal = {Journal of Biogeography},
  publisher = {Wiley},
  author = {Nekola,  Jeffrey C. and White,  Peter S.},
  year = {1999},
  month = jul,
  pages = {867–878}
}

@article{Spalding_2007,
  author={Spalding, Mark D. and Fox, Helen E. and Allen, Gerald R. and Davidson, Nick and Ferda{\~n}a, Zahra A. and Finlayson, Max A. X. and Robertson, J.},
  title={Marine ecoregions of the world: a bioregionalization of coastal and shelf areas},
  journal={BioScience},
  volume={57},
  number={7},
  pages={573--583},
  year={2007}
}

@article{Briggs_2012,
  author={Briggs, John C. and Bowen, Brian W.},
  title={A realignment of marine biogeographic provinces with particular reference to fish distributions},
  journal={Journal of Biogeography},
  volume={39},
  number={1},
  pages={12--30},
  year={2012}
}

@article{Costello_2017,
  author={Costello, Mark J. and Tsai, Po-Lynn and Wong, Pak-Sin and Cheung, Amy K. L. and Basher, Zahra and Chaudhary, Charan},
  title={Marine biogeographic realms and species endemicity},
  journal={Nature Communications},
  volume={8},
  number={1},
  pages={1057},
  year={2017}
}

@article{Longhurst_1995,
  author={Longhurst, Alan and Sathyendranath, Shubha and Platt, Trevor and Caverhill, Cathy},
  title={An estimate of global primary production in the ocean from satellite radiometer data},
  journal={Journal of Plankton Research},
  volume={17},
  number={6},
  pages={1245--1271},
  year={1995}
}

@article{Sathyendranath_1995,
  author={Sathyendranath, Shubha and Longhurst, Alan and Caverhill, Cathy M. and Platt, Trevor},
  title={Regionally and seasonally differentiated primary production in the North Atlantic},
  journal={Deep Sea Research Part I: Oceanographic Research Papers},
  volume={42},
  number={10},
  pages={1773--1802},
  year={1995}
}

@article{Hooker_2000,
  author={Hooker, Stanford B. and Rees, Nigel W. and Aiken, James},
  title={An objective methodology for identifying oceanic provinces},
  journal={Progress in Oceanography},
  volume={45},
  number={3-4},
  pages={313--338},
  year={2000}
}

@article{Oliver_2008,
  author={Oliver, Matthew J. and Irwin, Andrew J.},
  title={Objective global ocean biogeographic provinces},
  journal={Geophysical Research Letters},
  volume={35},
  number={15},
  year={2008}
}

@article{Reygondeau_2013,
  author={Reygondeau, Gabriel and Longhurst, Alan and Martinez, Emmanuel and Beaugrand, Gr{\'e}gory and Antoine, David and Maury, Olivier},
  title={Dynamic biogeochemical provinces in the global ocean},
  journal={Global Biogeochemical Cycles},
  volume={27},
  number={4},
  pages={1046--1058},
  year={2013}
}

@article{Elizondo_2021,
  author={Elizondo, Urdanibia H. and Righetti, D. and Benedetti, F. and Vogt, M.},
  title={Biome partitioning of the global ocean based on phytoplankton biogeography},
  journal={Progress in Oceanography},
  volume={194},
  pages={102530},
  year={2021}
}

@article{Sutton_2017,
  author={Sutton, Tracey T. and Clark, Malcolm R. and Dunn, Daniel C. and Halpin, Patrick N. and Rogers, Alex D. and Guinotte, John and Heino, M.},
  title={A global biogeographic classification of the mesopelagic zone},
  journal={Deep Sea Research Part I: Oceanographic Research Papers},
  volume={126},
  pages={85--102},
  year={2017}
}

@misc{Duarte_2022,
  author={Duarte, Carlos M.},
  title={Seafaring in the 21st century: the Malaspina 2010 circumnavigation expedition},
  year={2022}
}

@article{Sunagawa_2020,
  author={Sunagawa, Shinichi and others},
  title={Tara Oceans: towards global ocean ecosystems biology},
  journal={Nature Reviews Microbiology},
  volume={18},
  number={8},
  pages={428--445},
  year={2020}
}

@article{McNichol_2021,
  author={McNichol, Jesse and Berube, Paul M. and Biller, Steven J. and Fuhrman, Jed A.},
  title={Evaluating and improving small subunit rRNA PCR primer coverage for bacteria, archaea, and eukaryotes using metagenomes from global ocean surveys},
  journal={mSystems},
  volume={6},
  number={3},
  pages={10--1128},
  year={2021}
}

@incollection{Vincent_2022,
  author={Vincent, Fabrice and Ibarbalz, F. M. and Bowler, Chris},
  title={Global marine phytoplankton revealed by the Tara Oceans expedition},
  booktitle={Advances in Phytoplankton Ecology},
  pages={531--561},
  publisher={Elsevier},
  year={2022}
}

@article{Milke_2023,
  author={Milke, Felix and Meyerjürgens, Jan and Simon, Meinhard},
  title={Ecological mechanisms and current systems shape the modular structure of the global oceans’ prokaryotic seascape},
  journal={Nature Communications},
  volume={14},
  number={1},
  pages={6141},
  year={2023}
}

@article{Sokal_1958,
  author={Sokal, Robert R. and Michener, Charles D.},
  title={A statistical method for evaluating systematic relationships},
  journal={University of Kansas Science Bulletin},
  volume={38},
  number={22},
  pages={1409--1438},
  year={1958}
}

@article{mcnichol_characterizing_2025,
	title = {Characterizing organisms from three domains of life with universal primers from throughout the global ocean},
	volume = {12},
	issn = {2052-4463},
	url = {https://doi.org/10.1038/s41597-025-05423-9},
	doi = {10.1038/s41597-025-05423-9},
	number = {1},
	journal = {Scientific Data},
	author = {McNichol, Jesse and Williams, Nathan L. R. and Raut, Yubin and Carlson, Craig and Halewood, Elisa R. and Turk-Kubo, Kendra and Zehr, Jonathan P. and Rees, Andrew P. and Tarran, Glen and Gradoville, Mary R. and Wietz, Matthias and Bienhold, Christina and Metfies, Katja and Torres-Valdés, Sinhué and Mock, Thomas and Eggers, Sarah Lena and Jeffrey, Wade and Moss, Joseph and Berube, Paul and Biller, Steven and Bodrossy, Levente and Van De Kamp, Jodie and Brown, Mark and Sow, Swan L. S. and Armbrust, E. Virginia and Fuhrman, Jed},
	month = jul,
	year = {2025},
	pages = {1078},
}

@article{Raes2018,
  title = {Oceanographic boundaries constrain microbial diversity gradients in the South Pacific Ocean},
  volume = {115},
  ISSN = {1091-6490},
  url = {http://dx.doi.org/10.1073/pnas.1719335115},
  DOI = {10.1073/pnas.1719335115},
  number = {35},
  journal = {Proceedings of the National Academy of Sciences},
  publisher = {Proceedings of the National Academy of Sciences},
  author = {Raes,  Eric J. and Bodrossy,  Levente and van de Kamp,  Jodie and Bissett,  Andrew and Ostrowski,  Martin and Brown,  Mark V. and Sow,  Swan L. S. and Sloyan,  Bernadette and Waite,  Anya M.},
  year = {2018},
  month = aug 
}

@article{vanOostende2023,
  title = {Global ocean colour trends in biogeochemical provinces},
  volume = {10},
  ISSN = {2296-7745},
  url = {http://dx.doi.org/10.3389/fmars.2023.1052166},
  DOI = {10.3389/fmars.2023.1052166},
  journal = {Frontiers in Marine Science},
  publisher = {Frontiers Media SA},
  author = {van Oostende,  Marit and Hieronymi,  Martin and Krasemann,  Hajo and Baschek,  Burkard},
  year = {2023},
  month = may 
}

@article{Reygondeau2020,
  title = {Climate Change-Induced Emergence of Novel Biogeochemical Provinces},
  volume = {7},
  ISSN = {2296-7745},
  url = {http://dx.doi.org/10.3389/fmars.2020.00657},
  DOI = {10.3389/fmars.2020.00657},
  journal = {Frontiers in Marine Science},
  publisher = {Frontiers Media SA},
  author = {Reygondeau,  Gabriel and Cheung,  William W. L. and Wabnitz,  Colette C. C. and Lam,  Vicky W. Y. and Fr\"{o}licher,  Thomas and Maury,  Olivier},
  year = {2020},
  month = oct 
}

@article{Tagliabue2021,
  title = {Persistent Uncertainties in Ocean Net Primary Production Climate Change Projections at Regional Scales Raise Challenges for Assessing Impacts on Ecosystem Services},
  volume = {3},
  ISSN = {2624-9553},
  url = {http://dx.doi.org/10.3389/fclim.2021.738224},
  DOI = {10.3389/fclim.2021.738224},
  journal = {Frontiers in Climate},
  publisher = {Frontiers Media SA},
  author = {Tagliabue,  Alessandro and Kwiatkowski,  Lester and Bopp,  Laurent and Butensch\"{o}n,  Momme and Cheung,  William and Lengaigne,  Matthieu and Vialard,  Jerome},
  year = {2021},
  month = nov 
}

@article{Barber1996,
  author    = {Barber, R. T. and Sanderson, M. P. and Lindley, S. T. and Chai, F. and Newton, J. and Trees, C. C. and Chavez, F. P.},
  title     = {Primary productivity and its regulation in the equatorial Pacific during and following the 1991--1992 {E}l {N}i{\~n}o},
  journal   = {Deep Sea Research Part II: Topical Studies in Oceanography},
  volume    = {43},
  number    = {4-6},
  pages     = {933--969},
  year      = {1996},
}

@article{Carlson2022,
  title = {Viruses affect picocyanobacterial abundance and biogeography in the North Pacific Ocean},
  volume = {7},
  ISSN = {2058-5276},
  url = {http://dx.doi.org/10.1038/s41564-022-01088-x},
  DOI = {10.1038/s41564-022-01088-x},
  number = {4},
  journal = {Nature Microbiology},
  publisher = {Springer Science and Business Media LLC},
  author = {Carlson,  Michael. C. G. and Ribalet,  Fran\c{c}ois and Maidanik,  Ilia and Durham,  Bryndan P. and Hulata,  Yotam and Ferrón,  Sara and Weissenbach,  Julia and Shamir,  Nitzan and Goldin,  Svetlana and Baran,  Nava and Cael,  B. B. and Karl,  David M. and White,  Angelicque E. and Armbrust,  E. Virginia and Lindell,  Debbie},
  year = {2022},
  month = apr,
  pages = {570–580}
}

@article{Follett2022,
  title = {Trophic interactions with heterotrophic bacteria limit the range of
            Prochlorococcus},
  volume = {119},
  ISSN = {1091-6490},
  url = {http://dx.doi.org/10.1073/pnas.2110993118},
  DOI = {10.1073/pnas.2110993118},
  number = {2},
  journal = {Proceedings of the National Academy of Sciences},
  publisher = {Proceedings of the National Academy of Sciences},
  author = {Follett,  Christopher L. and Dutkiewicz,  Stephanie and Ribalet,  Fran\c{c}ois and Zakem,  Emily and Caron,  David and Armbrust,  E. Virginia and Follows,  Michael J.},
  year = {2022},
  month = jan 
}

@article{Morel2010,
  author  = {Andr\'e Morel and Herv\'e Claustre and Bernard Gentili},
  title   = {The most oligotrophic subtropical zones of the global ocean: similarities and differences in terms of chlorophyll and yellow substance},
  journal = {Biogeosciences},
  year    = {2010},
  volume  = {7},
  number  = {10},
  pages   = {3139--3151},
  doi     = {10.5194/bg-7-3139-2010} 
}

@article{Flombaum2013Present,
  author    = {Flombaum, Pedro and Gallegos, José L. and Gordillo, Rodolfo A. and Rincón, José and Zabala, Lina L. and Jiao, Nianzhi and Karl, David M. and Li, William K. W. and Lomas, Michael W. and Veneziano, Daniele and Vera, Carolina S. and Vrugt, Jasper A. and Martiny, Adam C.},
  title     = {{Present and future global distributions of the marine Cyanobacteria Prochlorococcus and Synechococcus}},
  journal   = {Proceedings of the National Academy of Sciences},
  year      = {2013},
  volume    = {110},
  number    = {24},
  pages     = {9824--9829},
  doi       = {10.1073/pnas.1307701110}
}

@article{cohen2021dinoflagellates,
  title={Dinoflagellates alter their carbon and nutrient metabolic strategies across environmental gradients in the central Pacific Ocean},
  author={Cohen, Natalie R and McIlvin, Matthew R and Moran, Dawn M and Held, Noelle A and Saunders, Jaclyn K and Hawco, Nicholas J and Brosnahan, Michael and DiTullio, Giacomo R and Lamborg, Carl and McCrow, John P and others},
  journal={Nature Microbiology},
  volume={6},
  number={2},
  pages={173--186},
  year={2021},
  publisher={Nature Publishing Group UK London}
}

@article{Dutkiewicz2024Multiple,
  title     = {Multiple biotic interactions establish phytoplankton community structure across environmental gradients},
  author    = {Dutkiewicz, Stephanie and Follett, Christopher L. and Follows, Michael J. and Henderikx-Freitas, Fernanda and Ribalet, Francois and Gradoville, Mary R. and Coesel, Sacha N. and Farnelid, Hanna and Finkel, Zoe V. and Irwin, Andrew J. and Jahn, Oliver and Karl, David M. and Mattern, Jann Paul and White, Angelicque E. and Zehr, Jonathan P. and Armbrust, E. Virginia},
  journal   = {Limnology and Oceanography},
  volume    = {69},
  issue     = {5},
  pages     = {1086--1100},
  year      = {2024},
  doi       = {10.1002/lno.12555}
}

@article{Barber1991Regulation,
  author    = {Barber, Richard T. and Chavez, Francisco P.},
  title     = {{Regulation of primary productivity rate in the equatorial Pacific}},
  journal   = {Limnology and Oceanography},
  year      = {1991},
  volume    = {36},
  number    = {8},
  pages     = {1803--1815},
}

@article{Dugdale1998Silicate,
  author    = {Dugdale, Richard C. and Wilkerson, Frances P.},
  title     = {{Silicate regulation of new production in the equatorial Pacific upwelling}},
  journal   = {Nature},
  year      = {1998},
  volume    = {391},
  number    = {6664},
  pages     = {270--273},
  doi       = {10.1038/34630},
}

@article{Vichi2011,
  title = {The emergence of ocean biogeochemical provinces: A quantitative assessment and a diagnostic for model evaluation: MODEL ASSESSMENT OF OCEAN PROVINCES},
  volume = {25},
  ISSN = {0886-6236},
  url = {http://dx.doi.org/10.1029/2010GB003867},
  DOI = {10.1029/2010gb003867},
  number = {2},
  journal = {Global Biogeochemical Cycles},
  publisher = {American Geophysical Union (AGU)},
  author = {Vichi,  Marcello and Allen,  J. Icarus and Masina,  Simona and Hardman-Mountford,  Nicholas J.},
  year = {2011},
  month = may
}

@article{Dorrell2012,
  title = {What makes a chloroplast? Reconstructing the establishment of photosynthetic symbioses},
  ISSN = {0021-9533},
  url = {http://dx.doi.org/10.1242/jcs.102285},
  DOI = {10.1242/jcs.102285},
  journal = {Journal of Cell Science},
  publisher = {The Company of Biologists},
  author = {Dorrell,  Richard G. and Howe,  Christopher J.},
  year = {2012},
  month = jan 
}

@article{LilaKoumandou2004,
  title = {Dinoflagellate chloroplasts – where have all the genes gone?},
  volume = {20},
  ISSN = {0168-9525},
  url = {http://dx.doi.org/10.1016/j.tig.2004.03.008},
  DOI = {10.1016/j.tig.2004.03.008},
  number = {5},
  journal = {Trends in Genetics},
  publisher = {Elsevier BV},
  author = {Lila Koumandou,  V. and Nisbet,  R.Ellen R. and Barbrook,  Adrian C. and Howe,  Christopher J.},
  year = {2004},
  month = may,
  pages = {261–267}
}

@article{falkowski1998biogeochemical,
  title={Biogeochemical controls and feedbacks on ocean primary production},
  author={Falkowski, Paul G and Barber, Richard T and Smetacek, Victor},
  journal={science},
  volume={281},
  number={5374},
  pages={200--206},
  year={1998},
  publisher={American Association for the Advancement of Science}
}

@article{field1998primary,
  title={Primary production of the biosphere: integrating terrestrial and oceanic components},
  author={Field, Christopher B and Behrenfeld, Michael J and Randerson, James T and Falkowski, Paul},
  journal={science},
  volume={281},
  number={5374},
  pages={237--240},
  year={1998},
  publisher={American Association for the Advancement of Science}
}

@article{johnson2006niche,
  title={Niche partitioning among Prochlorococcus ecotypes along ocean-scale environmental gradients},
  author={Johnson, Zackary I and Zinser, Erik R and Coe, Allison and McNulty, Nathan P and Woodward, E Malcolm S and Chisholm, Sallie W},
  journal={Science},
  volume={311},
  number={5768},
  pages={1737--1740},
  year={2006},
  publisher={American Association for the Advancement of Science}
}

@article{biller2015prochlorococcus,
  title={Prochlorococcus: the structure and function of collective diversity},
  author={Biller, Steven J and Berube, Paul M and Lindell, Debbie and Chisholm, Sallie W},
  journal={Nature Reviews Microbiology},
  volume={13},
  number={1},
  pages={13--27},
  year={2015},
  publisher={Nature Publishing Group UK London}
}

@article{horstmann2022microbial,
  title={Microbial diversity through an oceanographic lens: refining the concept of ocean provinces through trophic-level analysis and productivity-specific length scales},
  author={H{\"o}rstmann, Cora and Buttigieg, Pier Luigi and John, Uwe and Raes, Eric J and Wolf-Gladrow, Dieter and Bracher, Astrid and Waite, Anya M},
  journal={Environmental microbiology},
  volume={24},
  number={1},
  pages={404--419},
  year={2022},
  publisher={Wiley Online Library}
}

@article{sonnewald2020elucidating,
  title={Elucidating ecological complexity: Unsupervised learning determines global marine eco-provinces},
  author={Sonnewald, Maike and Dutkiewicz, Stephanie and Hill, Christopher and Forget, Gael},
  journal={Science advances},
  volume={6},
  number={22},
  pages={eaay4740},
  year={2020},
  publisher={American Association for the Advancement of Science}
}

@article{kavanaugh2014hierarchical,
  title={Hierarchical and dynamic seascapes: A quantitative framework for scaling pelagic biogeochemistry and ecology},
  author={Kavanaugh, Maria T and Hales, Burke and Saraceno, Martin and Spitz, Yvette H and White, Angelicque E and Letelier, Ricardo M},
  journal={Progress in Oceanography},
  volume={120},
  pages={291--304},
  year={2014},
  publisher={Elsevier}
}

@article{zuniga2021sinking,
  title={Sinking diatom assemblages as a key driver for deep carbon and silicon export in the Scotia Sea (Southern Ocean)},
  author={Z{\'u}{\~n}iga, Diana and Sanchez-Vidal, Anna and Flexas, Mar{\'\i}a del Mar and Carroll, Dustin and Rufino, Marta M and Spreen, Gunnar and Calafat, Antoni and Abrantes, F{\'a}tima},
  journal={Frontiers in Earth Science},
  volume={9},
  pages={579198},
  year={2021},
  publisher={Frontiers Media SA}
}

@article{smith2015vertical,
  title={Vertical mixing, critical depths, and phytoplankton growth in the Ross Sea},
  author={Smith Jr, Walker O and Jones, Randolph M},
  journal={ICES Journal of Marine Science},
  volume={72},
  number={6},
  pages={1952--1960},
  year={2015},
  publisher={Oxford University Press}
}
